\documentclass[12pt]{iopart}
\usepackage{iopams}
\usepackage{setstack}
\usepackage{graphicx}

\begin{document}

\topical{Electrostatics in soft matter}

\author{Ren\'e Messina}

\address{
Institut f\"ur Theoretische Physik II -   
Heinrich-Heine-Universit\"at  D\"usseldorf,
Universit\"atsstr. 1,
D-40225 D\"usseldorf,
Germany
}
\ead{\it messina@thphy.uni-duesseldorf.de}
\begin{abstract}
Recent progress in the understanding of the effect of electrostatics in soft matter
is presented.
A vast amount of materials contains ions ranging from the molecular scale 
(e.g., electrolyte) to the meso/macroscopic one (e.g., charged colloidal particles or polyelectrolytes).   
Their (micro)structure and physicochemical properties are especially dictated by
the famous and redoubtable long-ranged Coulomb interaction.
In particular theoretical and simulational aspects, including the 
experimental motivations, will be discussed. 
\end{abstract}

\newpage
\newpage

\section{\label{cha:Intro} Introduction}


Probably one of the most well known and understood ionic materials is 
sodium chloride (NaCl). In its {\it solid} form (i.e., NaCl cubic-like crystalline lattice),
the experimentally measured heat of vaporization (7.92 eV) can be deduced (within about 
$10\%$) from a straightforward lattice sum of the 
form\footnote[1]{
 The resulting energy in equation (\ref{eq:Madelung}) corresponds to the cohesive  energy per NaCl molecule. 
 An ion (either ${\rm Na^+}$ or ${\rm Cl^-}$) is placed at the origin and $\alpha_j=+,-$ 
 depending on the type of ion sitting at the lattice position $\vec r_j$.
 $e = 1.602 \times 10^{-19} {\rm C}$  stands for the usual elementary charge,  
 $\epsilon_0 = 8.854 \times 10^{-12} {\rm F/m}$ for
 the vacuum permittivity, and $a=2.81 {\rm \AA}$ for the NaCl lattice parameter.
 }  
%
\begin{eqnarray}
\label{eq:Madelung}  
E_M = \frac{e^2}{4 \pi \epsilon_0} \sum_{lattice} \frac{\alpha_j}{|\vec r_j|}
\simeq -1.747 \frac{e^2}{4 \pi \epsilon_0 a}
\end{eqnarray}
%
leading to the theoretical Madelung energy (here $E_M = -8.94{\rm eV}$) 
\cite{Madelung_PhysZ_1918,Ewald_AnnPhys_1921}.
This striking good agreement demonstrates that electrostatics is indeed 
the relevant ingredient governing our ionic crystal \cite{Feynman_Madelung_NaCl_Book_2006}.
In its {\it liquid} form, NaCl plays a fundamental role in soft matter, 
since it controls the degree of screening of the Coulomb interaction in all water based solutions. 
It is exactly this type of problem that 
this review will address: Electrostatics in soft matter.

Virtually all materials are more or less charged at the mesoscopic scale,
depending on the degree of the polarizability of the embedding solvent 
(or matrix) and the solute particles (e. g., colloidal particles, polymers, membranes. etc.).
The most well known example of polar solvent is evidently water which plays a crucial role in 
life, biological processes  as well as industrial applications.    
When the solute particles are polar too, they can then dissociate into charged particles 
(also called {\it macroions}) and ({\it microscopic}) counterions.
The counterion distribution near macroions turns out to be decisive for the surface 
properties of the latter. 

The pioneering works of Gouy and Chapman
\cite{Gouy_JPhys_1910,Chapman_PhilMag_1913}, 
realized almost one century ago, concern the counterion distribution near 
a planar charged interface. Applying the presently called Poisson-Boltzmann theory, 
they demonstrated that the counterion distribution profile decays algebraically 
as a function of the separation from the wall with a characteristic length that 
is inversely proportional to the surface charge density of the wall.  
Ten years later, Debye and H\"uckel \cite{DH_1923} accomplished a fundamental  
advance towards the understanding of screening.
This theory originally developed for electrolytes 
(i.e. a solution of microscopic cations and anions such as ${\rm Na^+}$ and ${\rm Cl^-}$)
and based on the linearization of the Poisson-Boltzmann equation 
is now widely used in plasma and solid state physics.\footnote[2]{
Note that a similar potential of interaction (so-called Yukawa potential) 
arises at the subatomistic scale to describe the cohesion of the nuclear matter.
Nonetheless,  in nuclear physics, the interpretation of this potential in terms of screening is 
not adequate. } 

Mean-field theories are appealing tools due to their intuitive and clear physical basis, 
and are robust theories as long as {\it electrostatic correlations} are not too important. 
In many practical situations (chromatin, polyelectrolyte multilayering, charged colloidal 
suspension, etc.) electrostatic correlations are strong enough to make mean-field theories
fail even on a qualitative level.
Two striking and natural consequences of electrostatic correlations, that can not be explained
by mean-field theories,  are charge reversal 
(also called overcharging) and like charge attraction:
(i) Overcharging concerns the situation where a macroion is locally covered by a cloud of counterions
whose global charge overcompensates that of the macroion so that the net charge 
(or effective charge) changes sign;
(ii) Like charge attraction is the counterintuitive effective attraction between two 
macroions carrying the same electric charge sign.

A colloidal suspension, the classical material of soft matter science, 
can crystallize via a strong enough mutual electrostatic repulsion.
An understanding of the resulting phase behavior necessitates approaches
where particle-particle correlations must obviously be taken into account.
This constitutes another example where approaches going beyond the mean-field 
level are required.

The present work examines the problem of electrostatics in soft matter 
systems using simple theoretical models and computer simulations.
The role of the little counterions is addressed in chapter \ref{cha:OC}. 
The relevance of excluded volume 
(i.e. the finite hard-core size of the constitutive ions) is discussed in chapter \ref{cha:ex_vol}. 
The problem of image charges as occurring near curved dielectric interfaces is  
presented in chapter \ref{cha:image}.
The basic physics in more complex processes such as 
polyelectrolyte adsorption and multilayering 
is elucidated in chapter \ref{cha:PE}.
Colloidal dispersions in strong confinement are presented in chapter \ref{cha:2D}.
Finally, a conclusion and possible outlooks are provided in chapter \ref{cha:conclu}.

\section{\label{cha:OC} Electrolyte at interfaces}


\subsection{\label{sec:PB} Foundations of electrostatic mean field theories in soft matter}

This part deals with the foundations of the electrostatic mean field theories in soft matter.
It is written on a pedagogical level such that the non-specialist reader 
should be in a position to easily capture the underlying physics.
Nonetheless, the expert will also certainly find some clarifying ideas in the forthcoming discussion.

\subsubsection{Poisson-Boltzmann theory}

\begin{figure}
\centering
\includegraphics[width = 10.0 cm]{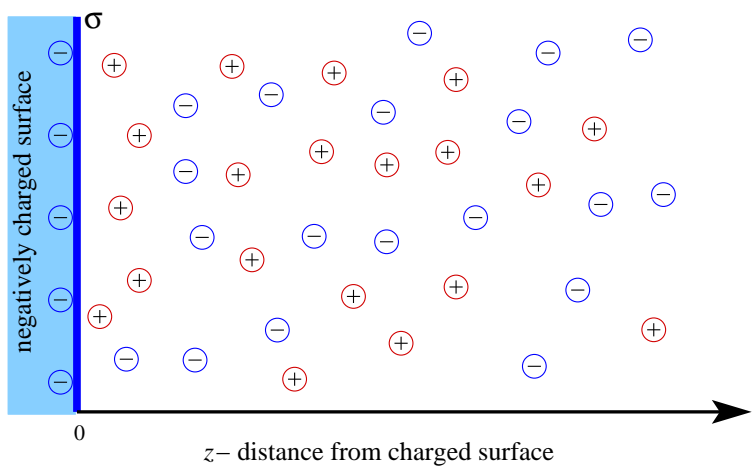}
\caption{
  Model for a simple electrolyte near a (negatively) charged surface.
 }
\label{fig:electrolyte}
\end{figure}

The model system we have here in mind is sketched in figure \ref{fig:electrolyte}.
We have to deal with a uniformly charged interface with a surface charge density
$\sigma$, separating the semi-infinite substrate from a simple electrolyte
[i.e. univalent cations ($+$) and anions ($-$)] and whose bulk density is $\rho_0$.
The system is globally electroneutral and the embedding solvent is merely characterized by
its dielectric constant. 
In this context, the first theoretical determination of counterion distribution for an inhomogeneous
fluid was realized by Gouy \cite{Gouy_JPhys_1910} 
and Chapman \cite{Chapman_PhilMag_1913} independently almost one century ago. 
This mean-field approach corresponding to the so-called Poisson-Boltzmann
theory is going to be now explained and discussed. 

A central quantity in the statistical mechanics of fluids is the {\it potential of mean force} (PMF). 
The latter corresponds to the potential stemming from the effective force
between two objects. The term ``effective'' means here a thermodynamical averaging
whose form is dependent on the ensemble (e.g., canonical, grand canonical) 
under consideration.
For the sake of simplicity we will consider the thermodynamical 
(i.e., macroscopic) limit where all ensembles are equivalent. 

%
%

A good starting point  is provided by the exact Poisson equation relating the 
the mean electrostatic potential (MEP), $\psi(z)$, to the PMF $w_{\alpha}(z)$
as follows:
%
\begin{eqnarray}
\label{eq:Poisson-Eq}  
\Delta \psi(z) = -\frac{e \rho_0}{\epsilon_0 \epsilon_{solv}} 
\left\{ \exp \left[ -\beta w_+(z) \right] - \exp \left[ -\beta w_-(z) \right] \right\},
\end{eqnarray}
%
where $\epsilon_{solv}$ is
the relative permittivity of the solvent (for water $\epsilon_{solv} \approx 80$),
$\beta \equiv 1/(k_BT)$ the reduced inverse temperature with 
$k_B$ being the Boltzmann constant and $T$ the absolute temperature.
The central approximation of the PB theory is to now set:
%
\begin{eqnarray}
\label{eq:PB-PMF-approx}  
w_{\pm}(z) \overset{(\rm PB)}{\simeq} \pm e \psi(z),
\end{eqnarray}
%
such that the exact Poisson equation (\ref{eq:Poisson-Eq}) becomes in the framework of the PB 
theory:
%
\begin{eqnarray}
\label{eq:PB-eq}  
\Delta \psi(z) = \frac{2 e \rho_0}{\epsilon} \sinh\left[ \beta e \psi(z) \right]
\quad (\epsilon \equiv \epsilon_0 \epsilon_{solv})
\end{eqnarray}
%
which is the well known PB equation.
The resulting MEP reads \cite{Andelamn_book_1995}:
\begin{eqnarray}
\label{eq:PB_wall_MEP} 
\psi(z)  =  - \frac{2k_BT}{e} 
              \ln \left[ 
               \frac{1+\gamma \exp(-\kappa z)} {1-\gamma \exp(-\kappa z)} 
              \right], 
\end{eqnarray}
where $\gamma$ is given by the positive root of:
\begin{eqnarray}
\label{eq:PB_gamma} 
\gamma^2 + (2 \kappa b) \gamma - 1 = 0
\quad {\rm so~that} \quad 0 \leq \gamma = -\kappa b + \sqrt{1 + (\kappa b)^2} <1,
\end{eqnarray}
We have introduced here in equation (\ref{eq:PB_wall_MEP}) and equation (\ref{eq:PB_gamma})
two important length scales, namely the screening length $\kappa^{-1}$:
\begin{eqnarray}
\label{eq:PB_kappa} 
\kappa^2 \equiv 8\pi \ell_B \rho_0
\end{eqnarray}
and the Gouy-Chapman length $b$
\begin{eqnarray}
\label{eq:PB_GouyChapLen} 
b \equiv \frac{e}{2 \pi  \ell_B |\sigma|},
\end{eqnarray}
where $\ell_B$ is another third relevant length in charged soft matter
known as the Bjerrum length\footnote[1]{
 The physical interpretation of the Bjerrum length is straightforward:
 It is the distance between two elementary charges $e$ that leads to an electrostatic 
 interaction equating $k_BT$.
 }
and reads:
\begin{eqnarray}
\label{eq:PB_BjerrumLen} 
\ell_B \equiv  \frac{e^2}{4\pi \epsilon  k_B T}.
\end{eqnarray}

%
\begin{figure}
\centering
\includegraphics[width = 10.0 cm]{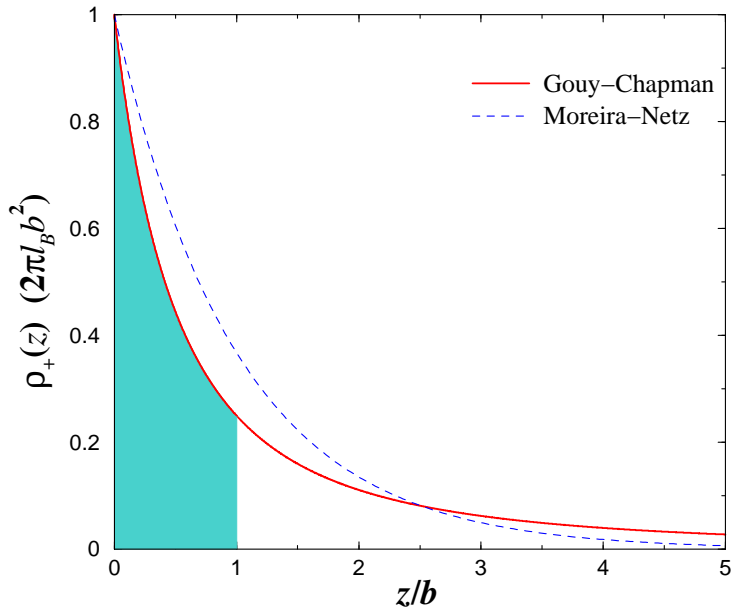}
\caption{
  Reduced (Gouy-Chapman) counterion distribution 
  $\rho_+(z) (2 \pi \ell_B b^2) =  \frac{1}{(1+z/b)^2}$
  as given by equation (\ref{eq:GC_wall_COU_distri}). 
  It is precisely at $z=b$, that the cumulated counterions (shadowed region) half-compensate
  the charge of the surface. In other words, the counterion integrated charge at
  $z=b$ is exactly $-\sigma/2$. 
  The strong Coulomb coupling limit (Moreira-Netz) 
  $ \rho_+(z) (2 \pi \ell_B b^2)  =  \exp(-z/b)$
  as given by equation~(\ref{eq:SCC_cou_dist_netz}) is also shown for direct comparison.
  }
\label{fig:Gouy-Chapman}
\end{figure}
%

The {\it salt-free} case can be actually easily obtained by considering $\kappa b \to 0$ in
equation (\ref{eq:PB_wall_MEP}) and (\ref{eq:PB_gamma}). 
Doing so we find: 
%
\begin{eqnarray}
\label{eq:GC_wall_MEP_COU} 
\lim_{\kappa b \to 0} \psi(z)  =  \frac{2k_BT}{e} \ln \left[ 1 + z/b \right] 
+ \frac{2k_BT}{e} \ln \frac{\kappa b}{2}.
\end{eqnarray}
The corresponding counterion distribution,  
$\rho_+(z) = \rho_0 \exp \left[ -\beta e \psi(z) \right]$, 
is then merely given by:
%
\begin{eqnarray}
\label{eq:GC_wall_COU_distri} 
\rho_+(z)  =  \frac{1}{2 \pi \ell_B} \frac{1}{(z+b)^2}.
\quad {\rm (salt-free)}
\end{eqnarray}
It is this formula (\ref{eq:GC_wall_COU_distri}) that is referred to as the 
Gouy-Chapman counterion distribution.
The corresponding plot can be found in figure \ref{fig:Gouy-Chapman}.

To better understand the physical meaning involved in the approximation 
(\ref{eq:PB-PMF-approx}), we shall make use of the  
exact so-called Yvon-Born-Green (YBG) hierarchy \cite{Hansen_BGY_Book_2006} that reads:
%
\begin{eqnarray}
\label{eq:BGY}  
- \vec \nabla_1 w_{\alpha}(z_1) = 
- \frac{q_{\alpha} |\sigma|}{2\epsilon} \vec e_z
- \sum_{\beta=1}^{2} 
\int \vec \nabla_1 \left[ \frac{q_{\alpha} q_{\beta}}{4\pi\epsilon r_{12}}  \right]
g_{\alpha \beta}(r_{12}, z_1, z_2) \rho_{\beta}(z_2) d^3r_2.
\end{eqnarray}
%
Equation (\ref{eq:BGY}) can be seen as a ``Newtonian'' version 
of the statistical Poisson equation (\ref{eq:Poisson-Eq}).
The left hand-side of equation (\ref{eq:BGY}) represents the effective force felt by the test ion 1 
of species $\alpha=\pm$ at prescribed location $\vec r_1 = (x_1,y_1,z_1)$.
The right hand side of equation (\ref{eq:BGY}) is made up of two contributions:
(i) The first term is merely the Coulomb interaction between the  charged interface and
the test ion 1. (ii) The second term involves the interaction between the test ion 1 
and the remaining solute ions, with $g_{\alpha \beta}(\vec r_1, \vec r_2)$ being
the pair distribution function and $\rho_{\beta}(z_2)$ the local ion density.
If the former is approximated by 
$g_{\alpha \beta}(\vec r_1, \vec r_2) \approx 1$ then equation (\ref{eq:BGY}) becomes:
%
\begin{eqnarray}
\label{eq:BGY_PB}  
- \vec \nabla_1 w_{\alpha}(z_1) =
- \vec \nabla_1  
\underbrace{ \left[ q_{\alpha} 
\left\{
\frac{|\sigma|}{2\epsilon} z_1
+ \sum_{\beta=1}^{2} 
\int \frac{q_{\beta}}{4\pi\epsilon r_{12}}  
\rho_{\beta}(z_2) d^3r_2
\right\}
\right]}_{= q_{\alpha} \psi(z_1) },
\end{eqnarray}
%
so that the potential of mean force reduces to the MEP times the charge,
which is precisely the PB approximation. In other words, the PB
theory neglects the  (lateral) {\it ion-ion correlations} in that sense
that   $g_{\alpha \beta}(\vec r_1, \vec r_2)=1$.\footnote[1]{
 Note that the existence of lateral ion-ion correlations 
 [i.e., $g_{\alpha \beta}(\vec r_1, \vec r_2) - 1 \neq 0$]
 have two physical origins: (i) Electrostatics 
 and (ii) steric effects due to excluded volume. 
 The latter are implicitly ignored in the PB framework.  
 } 
It is for this reason that the PB theory is a mean-field one.
Recalling that 
%
%
\begin{eqnarray}
\label{eq:g2_rho2}  
g_{\alpha \beta}(r_{12}, z_1, z_2)
\equiv
\frac{\rho_{\alpha \beta}^{(2)}(r_{12}, z_1, z_2)}{\rho_{\alpha}(z_1) \rho_{\beta}(z_2)},
\end{eqnarray}
%
where $\rho_{\alpha \beta}^{(2)}(r_{12}, z_1, z_2)$ is the two-particle density function,
one can equally well provide a geometrical interpretation:
The probability of finding two ions anywhere in the solution 
is {\it independent} of their relative  separation in the PB framework.\footnote[2]{
 Clearly, the bare Coulomb pair force between all constitutive ions are  
 properly taken into account in the PB theory, see equation (\ref{eq:BGY_PB}).
 It is the assumption of a structureless lateral arrangement of the ions 
 that creates the crucial inconsistency in the PB framework. 
 }
\subsubsection{Debye H\"uckel theory}  

In general the strongly non-linear PB equation (\ref{eq:PB-eq}) can not be solved analytically, and
its linearized version is therefore employed instead.
This latter approach was historically first developed by Debye and H\"uckel \cite{DH_1923}.
When the MEP is everywhere small (i.e., $|e\psi|<1$), the PB equation reduces to
%
\begin{eqnarray}
\label{eq:DH_eq} 
\Delta \psi
& = & 
\kappa^2 \psi
\end{eqnarray}
and the corresponding solution reads:
%
\begin{eqnarray}
\label{eq:DH_MEP} 
\psi(z) = \psi_S e^{-\kappa z} = - \frac{4\gamma k_BT}{e} e^{-\kappa z},
\end{eqnarray}
where $\psi_S$ denotes the surface potential. This result can be obtained either by
directly solving the DH equation (\ref{eq:DH_eq}) or substituting the small
$\psi_S$ value in the full PB solution equations (\ref{eq:PB_wall_MEP}) and (\ref{eq:PB_gamma}).


In order to more deeply understand the physical meaning of the
linear approximation, we shall rewrite the DH equation (\ref{eq:DH_eq}) in an
equivalent integral equation form.
In this context, it is instructive to use an approximative closure 
for the (exact) Orstein-Zernike equation,
as done by McQuarrie for a bulk electrolyte \cite{McQuarrie_DH_note_book_1976}, 
which leads to the DH description:
%
\begin{eqnarray}
\label{eq:DH_OZ_1} 
h_{0\alpha}( z_1) & = & 
\underbrace{c_{0 \alpha} (z_1)}_{\overset{\rm (DH)}{\approx} -\alpha z_1 / b} 
+ \sum_{\beta=1}^{2} \rho_{\beta} \int h_{0 \beta}(z_2) 
\underbrace{c_{\beta \alpha} (r_{12}, z_1, z_2) }_{\overset{\rm (DH)}{\approx} -\alpha  \beta \ell_B / r_{12}}
d^3r_2  \\
\label{eq:DH_OZ_2} 
& = &
\underbrace{c_{0 \alpha} (z_1)}_{\overset{\rm (DH)}{\approx} -\alpha z_1 / b} 
+ \sum_{\beta=1}^{2} \rho_{\beta} \int 
\underbrace{c_{0 \beta}(z_2)}_{\overset{\rm (DH)}{\approx} - \beta z_2 / b} 
h_{\beta \alpha} (r_{12}, z_1, z_2) d^3r_2,  
\end{eqnarray}
%
where, the subscript ``$0$'' in equations (\ref{eq:DH_OZ_1}) and (\ref{eq:DH_OZ_2}) 
stands for the charged interface (that can be envisioned as a special 
particle species at infinite dilution). Thereby, this notation preserves  
nicely the analogy with the bulk case.
The DH theory is readily obtained upon assuming that the direct correlation function
is equal to the (sign reversed) reduced pair potential 
[i.e., $c(r)=-\beta V(r)$], which becomes exact when $\rho_0 \ell_B^3 \to 0$ 
and $b/\ell_B \to \infty$.
In practice it is the first line (\ref{eq:DH_OZ_1}) that is used in  fluid theory to 
solve self-consistently the total correlation function $h_{ij} \equiv g_{ij}-1$
or the PMF via $h_{ij} \overset{\rm (DH)}{\approx} -\beta w_{ij}$.

For the sake of our discussion, however, it is the second line (\ref{eq:DH_OZ_2}) that turns out 
to be instructive. Indeed we see now that in the DH theory, the term
$g_{\beta \alpha} (\vec r_1, \vec r_2)$ is not trivially unity, 
since $h_{\beta \alpha} (\vec r_1, \vec r_2)$ does not vanish in equation (\ref{eq:DH_OZ_2}),
in contrast to what happens in the PB situation.
Hence ion-ion correlations are {\it not} neglected.\footnote[1]{
Note that MacQuarrie used the very same method [equation (\ref{eq:DH_OZ_2})] to determine 
analytically (via Fourier transformation) 
the DH potential in spherical geometry \cite{McQuarrie_DH_note_book_1976}.
However, in the past, he was not aware of the relevance of lateral ion-ion correlations,
and therefore did not point out this issue.
} 
This might seem at first sight counter-intuitive since the DH theory is based on the linearization of 
PB equation which ignores lateral correlations.
This being said, in the weak Coulomb regime where the DH theory is supposed to be valid, the 
deviations from the uncorrelated limit are then small. 

\subsection{\label{sec.SCC} Strong Coulomb coupling}

\subsubsection{\label{sec.SCC_netz_shklovskii} Strong Coulomb coupling theories}

This last decade \cite{Shklowskii_PRE_1999b,Moreira_EPL_2000}, 
a remarkable theoretical achievement has been accomplished
in the other extreme limit of strong Coulomb coupling.
More specifically, the counterion distribution near
a charged planar wall  has been predicted analytically and independently
by  Shklovskii \cite{Shklowskii_PRE_1999b} and 
Moreira and Netz \cite{Moreira_EPL_2000} in the strong Coulomb coupling regime
(i.e., the Gouy-Chapman problem at low temperature). 
A common and universal feature of these two works is
that the counterion distribution decays exponentially 
like $\exp(-z/b)$. 
These two approaches are going to be now briefly presented. 
%
\begin{itemize}
\item 
Using a field theoretic formulation applied to charged fluids 
\cite{Netz_EPL_1999,Netz_EPJE_2000}, Moreira and Netz \cite{Moreira_EPL_2000}
showed that at high Coulomb coupling 
(i.e.: for $\Xi \equiv \frac{\ell_B}{b} \gg 1$) the counterion distribution
obeys the following {\it exact} and elegant limiting law:
%
\begin{eqnarray}
\label{eq:SCC_cou_dist_netz} 
\frac{\rho(z)}{2 \pi \ell_B \sigma_s^2} = \exp(-z/b)  
\end{eqnarray}
%
with $\sigma_s= |\sigma|/e$ (having the dimension of the inverse of a surface)
standing for the number of elementary charges per unit area. 
A plot of equation~(\ref{eq:SCC_cou_dist_netz}) can be found in figure \ref{fig:Gouy-Chapman},
where a convenient visual comparison with the high temperature limit
[equation (\ref{eq:GC_wall_COU_distri})] 
is offered.
%
\item 
Using a fully different and more intuitive approach,  
Shklovskii \cite{Shklowskii_PRL_1999a} has applied Wigner crystal (WC) concepts 
\cite{Bonsall_PRB_1977,Rouzina_JCP_1996} 
to the problem of soft charged matter at effective low temperature.
Using some heuristic but physically sound arguments, 
essentially based on the simple fact that a ``desorbed'' counterion from 
the (triangular) WC counterion layer is correlated to the hole left behind
over the Gouy-Chapman length $b$,  
Shklovskii \cite{Shklowskii_PRL_1999a} obtains (up to the here important prefactor) 
the same result [equation (\ref{eq:SCC_cou_dist_netz})]  as Netz.
Interestingly, if one combines (i) the WC approach that provides the correct
exponential decay $\exp(-z/b)$ and (ii) the contact theorem 
which imposes the prefactor, 
$2 \pi \ell_B \sigma_s^2$ \cite{Wennerstroem_JCP_1982},\footnote[2]{
 Note that $2 \pi \ell_B \sigma_s^2 = \frac{1}{2\pi \ell_B b^2}$, 
 so that the PB theory predicts the exact contact value  as well
 [compare with equation (\ref{eq:GC_wall_COU_distri})], 
 see figure \ref{fig:Gouy-Chapman}. 
 This is not true, however, for the DH version.}
then one recovers the exact answer (\ref{eq:SCC_cou_dist_netz}). 

\end{itemize}
%

\subsubsection{Overcharging and Thomson problem}

As long as the Coulomb coupling between ions is ``fairly'' moderate
(which is the case for monovalent ions in aqueous solution), 
the PB theory \cite{Gouy_JPhys_1910,Chapman_PhilMag_1913,Manning_JCP_1969,LeBret_BioPol_1984a}
and even the DH one \cite{LevinFisher_PhysicaA_1995} 
describe astonishingly well the ion distribution (and hence the thermodynamical system properties) 
when compared to computer simulations 
\cite{Jonsson_JPC_1980,LeBret_BioPol_1984b,Deserno_Macromol_2000,Barbosa_PRE_2004,Linse_JCP_2005,Messina_EPL_2006}, 
theories going beyond the mean-field PB level
\cite{Belloni_CollSurf_1998},
and even experiments \cite{Bu_Langmuir_2004}. 
Nonetheless, as soon as ion-ion correlations get relevant, mean field theories
such as the PB one \cite{Trizac_PRE_1999} 
or its linearized version (as related above in \ref{sec:PB}) 
can not explain the experimentally observed relevant effect 
of overcharging \cite{Ruela_Macromol_2000,Lin_Langmuir_2004}.

%

\begin{figure}[t]
\centering
\includegraphics[width = 12.0 cm]{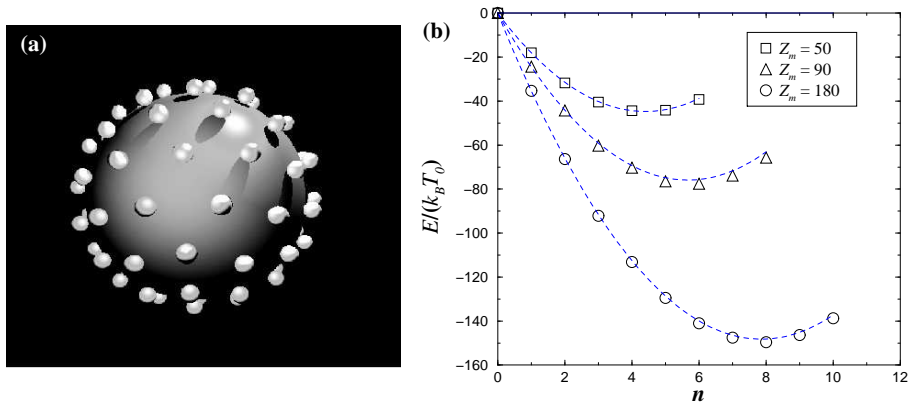}
\caption{
 (a)  
 Snapshot of the ground state structure  with $Z_m=180$ corresponding to $N=90$ counterions.
 Note the local triangular arrangement on the ``Thomson sphere''.
 (b)
 Electrostatic {\it ground state} energy (in units of $k_{B}T_{0}$ with $T_0$ being the room temperature) 
 as a function of the number of \textit{overcharging} counterions $n$ for three different bare charges $Z_m$. 
 The neutral case was chosen as the potential energy origin, and the curves were produced using the theory
 of equation (\ref{eq:OC_En_messina}), compare text.
 Data taken from \cite{Messina_PRE_2001}.
 } 
\label{fig:thomson_messina}
\end{figure}

Naively, one would think that the stable configuration corresponds to an exact neutralization
of the  macroion by the counterions.
This intuition is only correct for the case where the counterions are uniformly smeared 
out over the surface of the colloid. 
Indeed basic electrostatics show that,
for a central charge $Z_me<0$ (representing the macroion) and the shell of the counterions
of radius $a$ and (total) charge $Z_c^{(shell)}e>0$, 
the electrostatic potential energy is given by \cite{Messina_PRE_2001}
%
\begin{eqnarray}
\label{eq:OC_smeared_out_a} 
E=\frac{Z_mZ_c^{(shell)}e^2}{a}+\frac{{Z_c^{(shell)}}^2e^2}{2a}, \quad {\rm (CGS)}
\end{eqnarray}
%
where the first term describes the interaction between the central ion and 
the charged shell and the second one is the electrostatic energy stored in the shell
(i.e., the work done upon bringing the counterions from infinity to their current 
location $r=a$ of the shell).
Thereby, the criterion of stability
%
\begin{eqnarray}
\label{eq:OC_smeared_out_b} 
\frac{\partial E}{\partial Z_c^{(shell)}} = 0
~ {\rm and} ~ 
\frac{\partial^2 E}{\partial {Z_c^{(shell)}}^2} = e^2/a > 0
\quad 
\Rightarrow Z_c^{(shell)} =-Z_m
\end{eqnarray}
%
shows that the stable configuration corresponds to an exact neutralization.
In reality, the counterions are {\it discrete} and not smeared out, 
and when electrostatically bound to the macroion's surface,
they will maximize their separation such as to minimize the counterion-counterion repulsion.
%
This problem turns out to be exactly the one that was addressed one century ago by Thomson 
\cite{Thomson_PhilMag_1904} (also called the Thomson sphere or Thomson problem) 
who studied the ground state energy and structure
of $N$ (classical) electrons confined on a sphere (model of a classical atom).
The Thomson problem has only exact solutions for small $N$ 
and some magic numbers (e.g., $N=72$ corresponding to the fullerene structure) 
\cite{Wales_PRB_2006}.
Nonetheless, based on Wigner crystal ideas 
\cite{Bonsall_PRB_1977,Shklowskii_PRE_1999b,Grosberg_RevModPhys_2002}, an analytical model 
was developed which quantitatively accounts for the energy gain upon adsorbing 
overcharging counterions 
\footnote[1]{To achieve overcharging in nature one should normally add salt to the system
to ensure global electroneurality. For the sake of simplicity, however, we will
consider non-neutral systems because they can, on a very simple basis explain, why
colloids prefer to be overcharged.}
\cite{Messina_PRL_2000,Messina_PRE_2001}.
More precisely, the following relation for 
the energy variation $\Delta E_n$ 
(relative to the globally neutral state characterized by $n=0$ overcharging counterion 
and $N=Z_m/Z_c$ counterions, 
see figure \ref{fig:thomson_messina}  for a typical counterion arrangement) 
as a function of the number $n$ of (excess) 
overcharging $Z_c$-valent counterions \cite{Messina_PRE_2001} was derived:
%
\begin{eqnarray}
\label{eq:OC_En_messina} 
\Delta E_n = - \frac{\alpha Z_c^2}{\sqrt {4\pi a^2}} \left[ (N+n)^{3/2} - N^{3/2}  \right]
+ \frac{Z_c^2 n^2}{2a},
\quad
{\rm (CGS)}
\end{eqnarray}
%
where $\alpha$ ($\approx2$) is a numerical geometrical prefactor that was determined by simulations
(deduced from the value of $\Delta E_1$).
\footnote[2]{
Note that in the case of vanishing curvature 
(i.e., $a / d_c \to \infty$ where $d_c$ is the mean distance between counterions)
our expression becomes exact since the planar WC limit is recovered for which
$\alpha = 1.960 516...$ \cite{Bonsall_PRB_1977}.
} 
The first and attractive term in equation (\ref{eq:OC_En_messina}) stems basically
from the interaction between a counterion and its oppositely charged Wigner-Seitz cell. 
Energy profiles of equation (\ref{eq:OC_En_messina}) are sketched in figure \ref{fig:thomson_messina},
where one can see that these analytical predictions are pretty robust.
This simple approach to the understanding of the overcharging via the Thomson problem,
Wigner crystal concept and computer simulations has triggered a new interest in the community 
\cite{Levin_RepProgPhys_2002,Levin_EPL_2003,Patra_PRE_2003,Mukherjee_Langmuir_2003}
for the Thomson problem applied to soft matter.

%

We now consider the problem of a pair of macroions. In \cite{Messina_PRL_2000},
it was shown that two equally charged spheres are likely to be overcharged and undercharged
in the strong Coulomb coupling regime leading to a metastable {\it ionized state} that yields
a strong long-ranged attraction due to a {\it monopolar} contribution. 
All the mechanisms, so far reported in the literature, can only explain
{\it short-ranged} like-charge attraction
\cite{Jensen_Physica_1998,Allahyarov_PRL_1998,Linse_PRL_1999,
Netz_EPL_1999,Stevens_EPL_1990,Lau_PRL_2000,
Jensen_PRL_1997,Ha_PRL_1997,Shklowskii_PRL_1999a,
Kornyshev_PRL_1999,Arenzon_EPJB_1999,Naji_EPL_2004,Marcelo_PRL_2006}.

To further rationalize this phenomenon and the stability of ionized states
\cite{Messina_EPL_2000,Messina_PRE_2001}, 
two charged spheres of same radius $a$, carrying the same electric sign of charge 
but characterized by a charge ratio $\rho_Z$ such that $0<\rho_Z \equiv Z_B/Z_A\leq 1$,
were considered.    
Starting from a macroion pair where each macroion is neutralized by its counterions,
the process where a counterion is transfered from macroion $B$ (low bare charge) 
to macroion $A$ (high bare charge) was investigated \cite{Messina_EPL_2000,Messina_PRE_2001}. 
Having demonstrated that the ability of a macroion to get overcharged increases with growing 
(bare) surface charge density $\sigma$ (or the bare charge at fixed radius), it is clear that this
counterion-transfer process  will be energetically favorable below a certain value of $\rho_Z$.
This theoretical prediction shows that the criterion for stable ionized states 
(latter also called by other authors \cite{Patra_PRE_2003, Mukherjee_JPCM_2004} ``auto-ionization'') 
is governed by the value of
%
\begin{eqnarray}
\label{eq:Pauling_messina} 
\sqrt{N_A} - \sqrt{N_B} \gtrsim 1
\end{eqnarray}
%
(with $N_{A/B}=Z_{A/B}/Z_c$ being the number of counterions of macroion $A/B$) 
which reflects the correlation-hole energy difference between the two macroions 
(at identical radii). 
In particular, it was demonstrated that the higher the charge-asymmetry (i.e., $\rho_Z$) 
the more stable the ionized state and concomitantly the higher the degree of ionization
\cite{Messina_EPL_2000,Messina_PRE_2001}. 
The main findings related to this work \cite{Messina_PRL_2000,Messina_EPL_2000,Messina_PRE_2001}, 
can be summarized as follows:

\begin{itemize}
\item The ground state of a charged sphere is always overcharged due to counterion
      correlations.
\item At finite temperature and in the strong Coulomb regime 
      (accessible with multivalent {\it aqueous} ions), 
      colloids having different bare surface charge density auto-ionize due to counterion correlations. 
  
\end{itemize}

\subsection{\label{sec:dcc} Discretely charged surfaces}

\begin{figure}
\centering
\includegraphics[width = 12.0 cm]{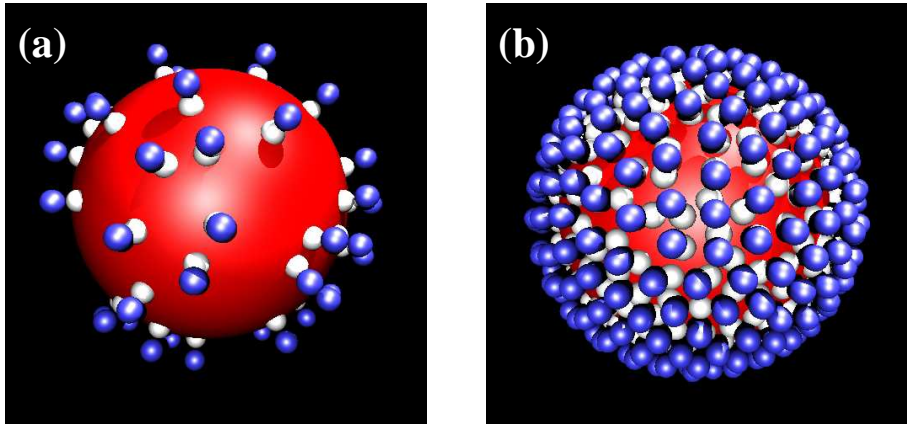}
\caption{
  Computer simulation snapshots of counterion ground state configurations.
  The discrete colloidal surface charges are in white. 
  The counterions are in blue.      
  (a) ``Low''  and (b) ``high''  surface charge densities are shown.
  Data taken from \cite{Messina_PhysicaA_2002}. 
 }
\label{fig:discrete_charges}
\end{figure}

The structural (i.e., bare) charge of spherical macroions is usually modeled
by a {\it central} charge, which, by virtue of the Gauss' law,  is equivalent to
a {\it uniformly} charged macroion surface as far as the electrostatic field (or potential)
{\it outside} the sphere is concerned.
However, in nature the charges on the colloidal surface are {\it discrete} 
(exactly as the counterions are) and localized, see figure \ref{fig:discrete_charges}. 
Thus, a natural question that rises is:
{\it Why} and {\it how} does the  counterion distribution 
depend on the way the structural charge of the macroion is represented 
(i.e., uniformly charged or discrete charges on its surface)?
It is precisely this problem that was addressed in  
\cite{Messina_EPJE_2001,Messina_PhysicaA_2002}.

%
%

{\it Why} is the counterion distribution sensitive to the choice of the representation
of the macroion charge (discrete vs. uniform)?
This question can be best answered by looking at and comparing the (intrinsic) 
electrostatic potentials generated by discretely and uniformly charged macroions 
(without counterions) \cite{Messina_EPJE_2001}. 
It was demonstrated in Ref. \cite{Messina_EPJE_2001} that the electrostatic potential at a reduced 
distance $r/a$ from the sphere (where $a$ stands for the distance of closest approach between an 
external unit test-charge and the macroion surface) may be significantly different according to the nature
of the macroion charge. 
In particular we show that the higher the bare surface charge 
(i.e., the closer we get to a uniform charge distribution) the shorter the correlation 
length (typically $r_c \sim \sqrt{1/\sigma_s}$) 
between the discrete surface charges, as intuitively expected.  
%
More specifically,
the contact potential is sensitive to the localization of the discrete charges, leading to
a pronounced depth in their vicinity.      
All those features, solely based on the spatial behavior of the electrostatic potential stemming
from the bare macroion, indicate that the counterion distribution should be much more complicated 
for a discrete macroion surface charge distribution than for the uniform case.

We now come to the other important question: 
{\it How} is the counterion distribution modified when introducing the more realistic discrete
macroion's surface charge distribution? 
This point is thoroughly addressed in \cite{Messina_PhysicaA_2002}, where two regimes
are considered: Ground state ($T=0$) and finite temperatures.
The corresponding relevant findings \cite{Messina_EPJE_2001,Messina_PhysicaA_2002} 
can be summed up as follows:
\begin{itemize}
\item At zero temperature, the counterion (surface) structure possesses greater order the
      the higher the reduced surface charge density $\sigma_s$ 
      and/or counterion valence $Z_c$ are.  

\item When overcharging comes into play several scenarios occur: 
      (i) At large $\sigma_s$, the overcharging is nearly the same as that obtained at
      a uniformly charged macroion's surface.
      (ii) At low $\sigma_s$ and for {\it monovalent} counterions, overcharging  
      is always weaker for discrete macroion charge distribution, due to the ion-pairing
      frustration for the excess counterions.
      (iii)  At low $\sigma_s$ and for {\it highly  multivalent} counterions, 
      overcharging can even be stronger in the discrete case due to ion-pairing.  

\item At finite temperature (in aqueous solutions), the volume counterion distribution
      is only affected for low $\sigma_s$ and  multivalent counterions.
\end{itemize}

The effect of surface charge discretization was later examined for 
different geometries by several groups 
\cite{Moreira_EPL_2002,Lukatsky_EPL_2002,Allahyarov_PRE_2003,
Henle_EPL_2004,Taboada-Serrano_JCP_2005,Qamhieh_JCP_2005,Madurga_JCP_2007}.

\section{\label{cha:ex_vol} The crucial role of excluded volume}

\subsection{Monovalent ions near a charged sphere}

So far, we have had a pretty good understanding of the physics involved in 
the counterion distribution for {\it salt-free} systems
where excluded volume effects are irrelevant.
The situation becomes much more complicated at finite salt-concentrations in aqueous solutions 
(i.e., water at room temperature in the presence of added salt),
where the Coulomb coupling is (rather) weak especially for monovalent ions.
Thereby, a direct application of Wigner crystal ideas is not straightforward enough
to account for the unexpected overcharging at {\it weak} Coulomb coupling
that was reported theoretically 
\cite{Spitzer_JColIntfSci_1983,Marcelo_JCP_1985,Kjellander_JCP_1998,Deserno_JBCB_2001}), 
but unexplained,  for {\it monovalent} salt-ions of large size. 

Molecular dynamics computer simulations as well as integral-equation theory 
\cite{Messina_EPL_2002} were employed to  identify the mechanisms that govern
counterions ordering and overcharging in this weak Coulomb coupling regime.  
Those mechanisms are as follows:
\begin{itemize}
\item Increasing the electrolyte particle {\it size} (at given salt concentration)
      decreases the available volume of the fluid (or equivalently its entropy) which 
      {\it favors ion-ion correlations}.
\item The interface provided by the macroion causes an increase of the ion density
      close to it, and concomitantly enhances the {\it lateral ordering} 
      (similar to the prefreezing phenomenon in {\it neutral} inhomogeneous fluids). 
\item Surface lateral ordering and (weak) Coulomb coupling lead to overcharging.
\end{itemize}

\subsection{Macroion adsorption at planar substrates}

\begin{figure}
\centering
\includegraphics[width = 10.0 cm]{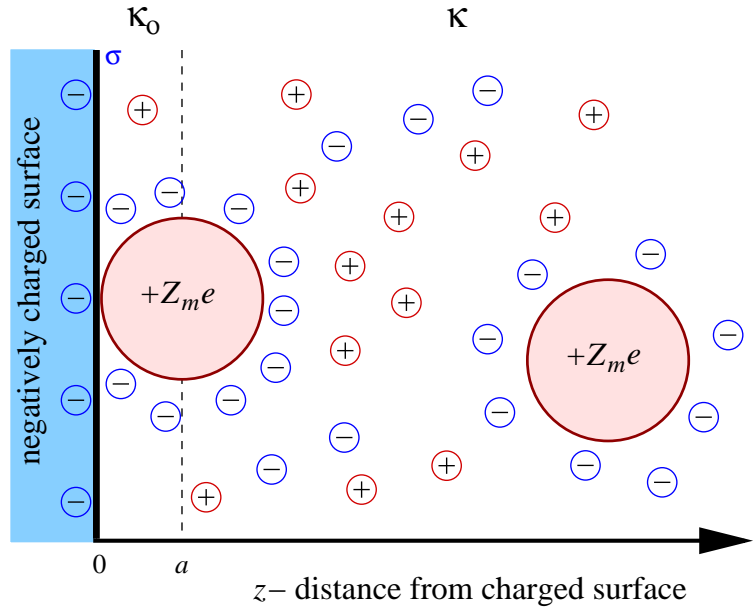}
\caption{
  Schematic view of the electrostatic model for macroions near an oppositely charged interface.
  The macroions are characterized by a distance of closest approach $z=a$ to the charged 
  surface, leading to two screening strengths $\kappa_0$ and $\kappa$ for $0<z<a$ and
  $z>a$, respectively.
  }
\label{fig:macroion_ads_model}
\end{figure}

Excluded volume effects coupled to electrostatic interactions can also lead to counter-intuitive
phenomena in the process of macroion adsorption. 
A description of the model setup is sketched in figure \ref{fig:macroion_ads_model}.
For instance, Jimenez-{\'A}ngeles and Lozada-Cassou showed,     
using integral equation theory, that for moderately 
(attractive) charged substrates, a film of {\it coions} first builds up.
The electrostatic consequence is that at the direct vicinity  of the surface of the substrate 
its charge gets amplified (i.e., {\it surface charge amplification}).
The driving force of this effect is due to the macroion-ion attractive correlations.
\footnote[1]{
The negative counterions of the positively charged macroions correspond to the 
coions of the planar substrate. 
Thereby, electrostatic correlations tend to localize the counterions of the macroions
over {\it its whole surface} in a uniform manner.
Hence, as long as the strength of the surface charge density of the oppositely charged
substrate is low enough, a finite number of counterions of the macroions should    
stay in the vicinity of the interface (see figure \ref{fig:macroion_ads_model}),
leading to a surface charge amplification. 
}
This effect was overlooked in the past, because the authors neglected either the finite size 
of the macroion  \cite{Netz_PRE_1999} 
or the spatial distribution of the little salt-ions \cite{GonzalesMozuelos_JCP_1991}.

%
%

Recently, this problem was revisited by using a very simple analytical model based on
the Debye-H{\"u}ckel approximation {\it but} taking into account the finite
size of the macroion via its distance of closest approach $a$ (i.e., its radius) 
to the wall (see figure \ref{fig:macroion_ads_model} as well) \cite{Messina_JCP_2007}. 
Two regimes were specifically examined: The strong and weak screening regimes which are now
briefly described.

\begin{itemize}
\item 
In the strong screening regime ($\kappa_0 a \gg 1$)
\footnote[1]{ 
$\kappa_0$ stands for the screening strength stemming uniquely from the little ions,
see also figure \ref{fig:macroion_ads_model}.
}
the wall-macroion attractive interaction is exclusively governed by the screening contrast 
$\kappa_0/\kappa$.  
\footnote[2]{ 
$\kappa$ stands for the total screening strength stemming  from all the ions
present in  the solution (also including the macroions),
see also figure \ref{fig:macroion_ads_model}.
}
More precisely, it was shown that the contact potential of interaction 
$U_m$ is merely given by \cite{Messina_JCP_2007}:
%
\begin{eqnarray}
\label{eq:high_ka_macro_ads} 
\beta U_m \simeq 1 - \frac{\kappa}{\kappa_0}.
\end{eqnarray}
%
\item
In the weak screening regime ($\kappa_0 a \ll 1$) and
for sufficiently small surface charge density
($\frac{\kappa b}{2Z_m} \gg 1$), the reduced electric field \footnote[3]{ 
The reduced electric field is defined as $E^*(z) \equiv -(b/2) e \frac{d\psi(z)}{dz}$
such that at the interface $z=0$ we have $E^*(0) = -1$.
}
at contact follows this simple law \cite{Messina_JCP_2007}:
%
\begin{eqnarray}
\label{eq:small_ka_macro_ads} 
E^*(a) \simeq - \frac{\kappa b}{2Z_m} \left( 1 - \frac{\kappa_0^2}{\kappa^2} \right) \kappa a.
\end{eqnarray}
%
This equation (\ref{eq:small_ka_macro_ads}) tells us that surface charge amplification
is increasing with growing colloidal particle size $a$ and increasing
Gouy-Chapman length $b$ (i.e., {\it decreasing} $\sigma_s$). 

\end{itemize}

\section{\label{cha:image}Image charges in spherical geometry}

In a typical experimental setup, the dielectric constant of a macroion 
is rather low ($\varepsilon_{m} \approx 2-5$) which is much smaller than that
of its embedding solvent (e. g., for water $\varepsilon_{solv} \approx 80$) leading to
a {\it high dielectric contrast}, 
$\Delta_{\varepsilon} \equiv \frac{\varepsilon_{solv} -\varepsilon_{m}}{\varepsilon_{solv}+\varepsilon_{m}}$, 
at the interface.  
It turns out that for a perfect {\it planar} substrate
(which can be envisioned as a colloid of vanishing curvature), 
there is an elegant analytical solution for the electric field.
More precisely, the electric field generated by  the induced surface charge at the interface positioned at
$z=0$ (due to the presence of a point-like ion of charge $q$ located at $z=\ell$) 
can be exactly obtained by a ``fictive'' point-like charge $q_{im}=\Delta_{\varepsilon}q$ 
located at the mirror position $z=-\ell$ \cite{Jackson_book_1975}. 
This feature corresponds to the so-called method of {\it image charges}.
The inclusion of such image forces for the case of an electrolyte close to a planar
dielectric interface was studied in the past by computer simulations
\cite{Torrie_JCP_1982,Torrie_JCP_1984,Bratko_CPL_86,Moreira_EPL_2002}, 
integral equation formalisms \cite{Kjellander_CPL_84,Kjellander_JCP_85}, 
mean-field \cite{Outhwaite_JCSFT_83,Netz_PRE_1999,Gruenberg_JPhysCondM_2000,Sens_PRL_2000,Bhuiyan_MolPhys_2007} 
and strong-coupling  \cite{Moreira_EPL_2002} theories.
As far as the cylindrical case 
\cite{Fixman_Macromol_1978,Aaron_JCP_2006,Aaron_JCP_2007}
is concerned, there is no simple ``image charge'' picture.

The problem of the dielectric discontinuity in {\it spherical} geometry is, already at the level
of a single ion interacting with a dielectric (neutral) sphere, 
considerably more complicated than its planar counterpart. 
Indeed, if we want to reformulate the problem  in terms of image charges,
one would need an {\it infinite}  number of image charges, thus 
making its usage much less attractive than in the planar case.
Due to this difficulty, the problem of image charges in spherical geometry is sparsely
studied in soft matter. 
Nevertheless, twenty years ago, Linse studied the counterion distribution with image forces 
around spherical charged micelles by means of Monte Carlo simulations \cite{Linse_JPC_1986}. 
In his work \cite{Linse_JPC_1986}, Linse used a {\it two-image} charge approximation instead of
the full continuous image charge distribution. The conclusions of his study 
remain qualitatively correct. 
The dielectric response of a dipolar fluid confined to a spherical 
cavity was recently addressed by Blaak and Hansen using MD simulations
\cite{Blaak_JCP_2006}.

In the field of image forces in spherical geometry, exact results for the electrostatics
of an ion interacting with a dielectric sphere 
(see figure \ref{fig:image_sphere} for the model geometry) 
were reported \cite{Messina_image_2002}. 
Furthermore, Monte Carlo simulations were performed
to elucidate the behavior of an electrolyte near a spherical macroion at finite dielectric contrast,
where image forces are properly taken into account \cite{Messina_image_2002}.
The main results are as follows \cite{Messina_image_2002}:
%
\begin{figure}
\centering
\includegraphics[width = 14.0 cm]{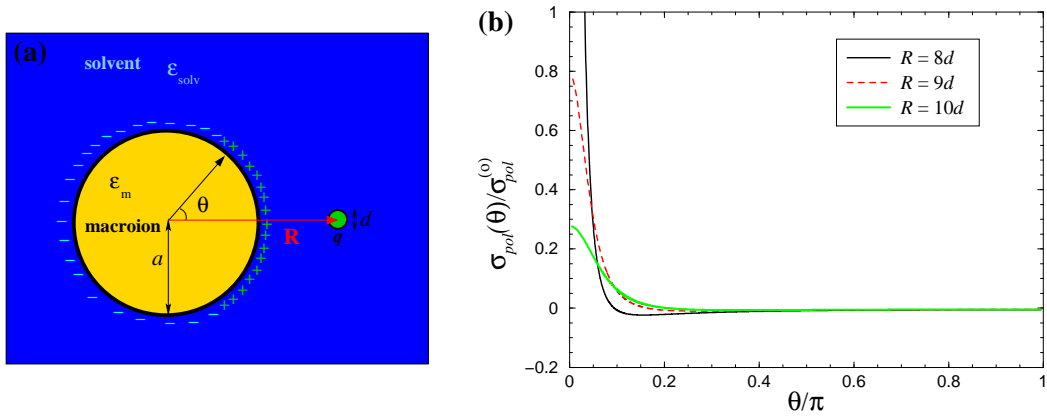}
\caption{
 (a)
  Model for a dielectric sphere (colloid) of dielectric constant $\varepsilon_m$
  embedded in an infinite medium (the solvent) characterized by a different dielectric
  constant $\varepsilon_{solv}$.
  A test positive charge ($q$) is located near the boundary outside the spherical particle at a radial distance $R$.
  The resulting induced surface polarization charges are also illustrated for the case
  where $\varepsilon_{solv}>\varepsilon_{m}$.
  Note that the global induced net charge vanishes.
  This is a two-dimensional representation of the three-dimensional system.
 (b)
  Polar profiles, as obtained from equation (\ref{eq:img_pol_charge}), 
  of the surface density of polarization charge
  $\sigma^{(sph)} _{pol}(\theta)$ in units of
  $\sigma ^{(0)}_{pol}=\frac{q}{4\pi \varepsilon_{solv} d^2}$
  for different radial distances $R$ of the test  charge $q$ with 
  $\varepsilon_{solv}=80$, $\varepsilon_m=2 $ and $a=7.5 d$.
 }
\label{fig:image_sphere}
\end{figure}
%
\begin{itemize}
\item {\it Single ion}: 
      A compact and exact analytical expression has been derived for the polar profile of
      the induced surface charge, and it reads:
      \begin{eqnarray}
      \label{eq:img_pol_charge}
      \sigma_{pol}(\theta ) & = &
      \frac{q (\varepsilon_{solv}-\varepsilon_m)}{4\pi \varepsilon_{solv} R^{2}}
      \sum ^{\infty }_{l=1}
      \left( \frac{a}{R}\right) ^{l-1}
      \frac{(2l+1)l}{\varepsilon_{solv}(l+1)+\varepsilon_ml}
      P_{l}(\cos \theta ),
      \end{eqnarray}
      %
      where $q$ is a test ion at a radial distance $R$ (see figure \ref{fig:image_sphere}) 
      and $P_l$ designates the Legendre polynomials of order $l$.
      The {\it strength} as well as the {\it range} of image forces in spherical geometry are 
      considerably smaller than at vanishing curvature, due to {\it auto-screening}.
\item {\it Electrolyte}: 
       For aqueous monovalent ions the (effective) image force is basically 
       equal to the {\it self-image} one (i.e., the interaction between an ion and its own image).
       However, when dealing with multivalent counterions, the {\it lateral} image-counterion 
       correlations can significantly affect the (local) counterion density and, as a major effect,
       they {\it screen} the self-image repulsion. 
       Upon adding salt, it was shown that the strength of the image forces induced 
       by the {\it coions} is marginal. Besides, overcharging is robust against image forces.
                    
\end{itemize}

Very recently, Re\v{s}\v{c}i\v{c} and Linse extended \cite{Linse_JCP_2008} 
the one-colloid problem to the
two-colloid interaction problem with dielectric discontinuity.
Using a cylindrical cell model and MC simulations, they found
(i) weaker counterion accumulation at the macroion's surfaces, 
(ii) stronger effective repulsion at moderate Coulomb coupling, and
(iii) a less attractive effective force at strong Coulomb coupling.
These findings are fully consistent with the one-colloid features just discussed above.

\section{\label{cha:PE}
Polyelectrolyte adsorption and multilayers}

Polyelectrolytes (PEs) are polymers containing a variable (usually large) amount of 
ionizable monomer along the chemical backbone. 
Once dissolved in a suitable polar solvent such as water, the ion pairs dissociate
by creating a charged chain with floating counterions.
PEs represent a broad and interesting class of materials that have attracted an increasing attention
in the scientific community. 
PEs have applications in modern technology as well as biology, since virtually all proteins, 
as well as DNA, are charged.
The adsorption of PEs onto surfaces is an important process, since they modify the
physico-chemical properties of the surface.
From a theoretical point of view, charged polymers (in bulk or adsorbed) are much less understood 
\cite{Barrat_Joanny_1996,Joanny_Houches_2001} than neutral ones \cite{DeGennes_Book_1979}. 
One of the main difficulties is the addition of new length scales
set by the tremendous long-ranged Coulomb interaction. 
Hence, the study of adsorption of PEs is motivated by fundamental aspects as well as practical ones.

\begin{figure}
\centering
\includegraphics[width = 7.0 cm]{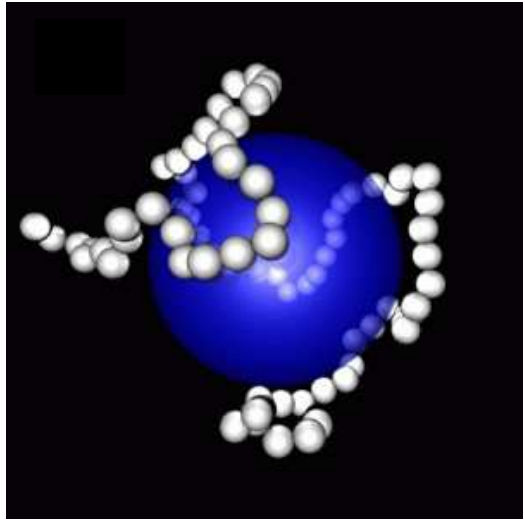}
\caption{
A computer simulation snapshot of PE-colloid complexation (tennis ball-like conformation)
\cite{Messina_Langmuir_2003}.
}
\label{fig.tennis_ball}
\end{figure}
%

\subsection{Polyelectrolyte-colloid complexation}

\subsubsection{Oppositely charged spherical substrates}

The works related to the interaction 
between PEs and  {\it oppositely} charged
spheres are here briefly reviewed.\footnote[1]{
The reader can also look at recent reviews 
\cite{Vries_CurrOpinCollIntSc_2006,Messina_review_2004}
on this field for more details. 
} 
The complexation of flexible PEs with oppositely charged macroions is a relevant
process in biology \cite{Schiessel_Review_2003}. For instance a nucleosome can be 
seen as an electrostatic binding between DNA and histone proteins,
where the latter can be envisioned as charged spheres.\footnote[2]{
We are aware that this assumption is at best a caricature of a real system 
(provided that non-specific interactions are dominant).
Nonetheless, from an electrostatic viewpoint, we think that the qualitative
features should be captured.}
Many theoreticians \cite{Muthukumar_JCP_1994,Sens_PRL_1999,
Mateescu_EPL_1999,Park_EPL_1999,Netz_Macromol_1999a,Kunze_PRL_2000,
Nguyen_Physica_2001,Schiessel_Macromol_2003,Vries_CurrOpinCollIntSc_2006,
Winkler_PRL_2006,Winkler_JCP_2006,Winkler_JPC_2007}
have investigated these types of objects to understand the electrostatics
governing those structures. Two very relevant results are: 
(i) the possible overcharging of the sphere by the long PE and (ii)
a strong wrapping of the PE about the sphere (see figure \ref{fig.tennis_ball} for an example).
A considerable effort was also provided by the simulators
\cite{Wallin_Langmuir_1996_I,Wallin_JPhysChem_1996_II,Wallin_JPhysChemB_1997_III,
Kong_JCP_1998,Jonsson_JCP_2001a,Jonsson_JCP_2001b,
Stoll_Macromolecules_2001,Stoll_JCP_2001,Stoll_Macromolecules_2002,
Akinchina_Macromolecules_2002,Dzubiella_Macromolecules_2003,Messina_Langmuir_2003,
Maiti_NanoLett_2006}
these last twelve years.
Some relevant findings in this field can be summarized as follows:
%
\begin{itemize}
\item 
{\it The effect of chain stiffness}, which was first systematically studied
by Wallin and Linse \cite{Wallin_Langmuir_1996_I} by MC simulations,
is an important key controlling the PE adsorption.
They showed that the lower the chain stiffness, the higher the 
PE adsorption and, concomitantly, the overcharging of the charged sphere 
by the PE.
Stoll and Chodanowski \cite{Stoll_Macromolecules_2002}, 
using MC simulation as well but with Yukawa potentials,
showed that upon increasing the chain stiffness, {\it solenoid}  conformations
are obtained as predicted analytically by Nguyen and Shklovskii \cite{Nguyen_Physica_2001}. 

\item 
{\it The effect of chain length} was also  addressed 
(by means of computer simulations) in the past 
\cite{Wallin_JPhysChemB_1997_III,Stoll_Macromolecules_2001,Stoll_JCP_2001,Akinchina_Macromolecules_2002}.
For large chain/sphere size ratio,   
Chodanowski and Stoll \cite{Stoll_Macromolecules_2001} found, for fully flexible chains,
that only a marginal portion of the PE gets adsorbed to the sphere, 
and the rest of the chain consists of extended tails.
At ``moderate'' chain/sphere size ratio \cite{Stoll_Macromolecules_2001},  
they found a strong PE collapse into a tennis-ball like structure
(as illustrated in figure \ref{fig.tennis_ball}).
Considering {\it both} the effects of chain length and the chain stiffness,
Akinchina and Linse \cite{Akinchina_Macromolecules_2002}
reported a rich phase behavior: Tennis-ball like, solenoid, 
multiloop (also called rosette \cite{Schiessel_Review_2003}),
single loop, as well as ``U''-shaped conformations.
Note, that there is remarkable agreement with the rosette structure
found theoretically by Schiessel et al. \cite{Schiessel_EPL_2000}.

\item 
The effect of the {\it discrete} nature of the protein charge distribution
was addressed by Carlsson et al. \cite{Carlsson_JPCB_2001}.
In their MC simulations \cite{Carlsson_JPCB_2001}, they found
that complexation can be stronger with a discrete protein charge distribution
(in agreement with the ideas discussed in \ref{sec:dcc}).

\item
{\it Multisphere} complexation involving many charged spheres bridged via oppositely 
charged PEs were investigated by Jonsson and Linse \cite{Jonsson_JCP_2001a,Jonsson_JCP_2001b}
by means of MC simulations. The effect of linear chain charge density, chain length, 
and macroion charge valency was addressed in Ref. \cite{Jonsson_JCP_2001a}.
Interestingly, at prescribed PE linear charge density, the authors found that complexation gets stronger
upon increasing the chain length \cite{Jonsson_JCP_2001a}.
The effect of chain flexibility was studied in \cite{Jonsson_JCP_2001b}, 
and  it was found that the macroion arrangement gradually becomes more linear and ordered
along the (long) chain when its stiffness is increased.  

\end{itemize}

\subsubsection{Like-charge complexation}

Whereas many studies have been devoted for the case of chain-sphere complexation
where the two charged bodies are oppositely charged, as we just saw, 
much less is known concerning the problem of {\it like-charge sphere-PE complexation}.      

In \cite{Messina_PRE_2002,Messina_JCP_2002}, the complexation between
a sphere and a long flexible PE ({\it both negatively charged}) was discussed.
Whereas like-charge attraction in the strong Coulomb coupling limit is expected 
(and therefore complexation too),
new and rather unexpected chain conformations are reported.   
Different coupling regimes as well as the influence of the 
linear charge density, $f$, of the PE chain were considered in 
\cite{Messina_JCP_2002}.
The relevant conclusions are as follows:
\begin{itemize}
\item At strong coupling the PE chain is always adsorbed in a {\it flat} structure,
      whose conformation strongly depends on $f$. At high $f$, the conformation consists
      of  densely packed monomers following a Hamiltonian-walk. Upon reducing $f$ the
      chain tends to spread more and more over the particle surface.
      These findings could have some relevance for organic solutions.     
\item Under {\it aqueous} conditions, complexation can be obtained with multivalent
      counterions and for high enough values of $f$. In contrast to the strong coupling 
      case, the formation of {\it loops} is reported.    
\end{itemize}

\subsection{Polyelectrolyte adsorption at planar surfaces}

The reader who wants to know a detailed account of the field 
of PE adsorption at surfaces is invited to consult  the recent reviews of 
Netz and Andelamn \cite{Netz_PhysRepRev_2003}
and of Dobrynin and Rubinstein \cite{Dobrynin_Review_2005}.   
In this part, one would like to propose some basic ideas and features supported
by MC simulations about the adsorption of highly charged polyelectrolytes onto oppositely charged {\it planar} 
surfaces in a salt-free environment \cite{Messina_PRE_2004,Messina_PRE_err_2006,Messina_JCP_2006}. 
Flexible \cite{Messina_PRE_2004,Messina_PRE_err_2006} 
as well as rod-like \cite{Messina_JCP_2006}  PEs are now discussed.

\subsubsection{\label{sec:PE_ads_entropy} Role of entropy}
There is a simple and clear entropic mechanism that influences 
{\it multi-polymer-chain} adsorption that is going to be pointed out first.
It can be best understood by recalling the {\it counterion release} 
effect:
The  adsorption process of {\it one} polyion of valence $Z$ 
typically leads to the release into solution of $Z$ 
(initially adsorbed) surface monovalent counterions, which is ``electrostatically invariant'' 
but entropically (highly) favorable. 
%
{\it This very same effect is also the reason why longer chains
can better adsorb at a prescribed monomer density}.
Indeed, at prescribed monomer density, increasing the chain length $N_m$\footnote[1]{  
 Rigorously, $N_m$ represents the number of monomers per chain 
 corresponding experimentally to the polymerization degree.}
involves decreasing the the number of chains. Thereby, the resulting (bulk) entropy
stemming from the PE chains becomes reduced accordingly. 
This entropic mechanism linked to the chain length at prescribed monomer density 
is henceforth referred to as: 
{\it polymerization induced adsorption}.

\subsubsection{Flexible chains} \cite{Messina_PRE_2004}

\begin{figure}
\centering
\includegraphics[width = 12.0 cm]{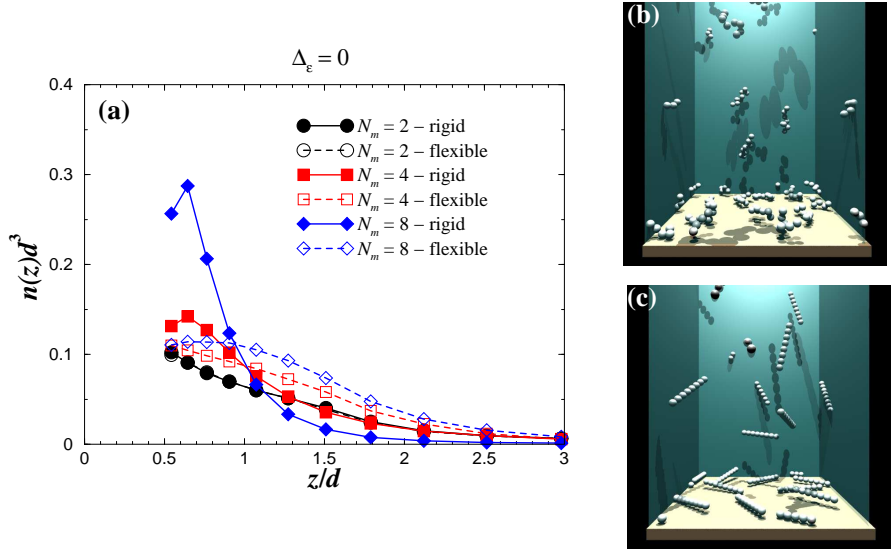}
\caption{
(a) Profiles of the monomer density $n(z)$ for various chain length $N_m$: 
{\it flexible} vs {\it rigid} chains \cite{Messina_JCP_2006}. 
Snapshots at $N_m=8$ for (b) flexible and (c) rigid chains.
}
\label{fig.PE_rig_flex_nz}
\end{figure}

When {\it no} image forces are present (i.e., $\Delta_{\epsilon}=0$), 
it was found that the monomer density profile, $n(z)$, decays monotonically 
for very short chains even near contact, see figure \ref{fig.PE_rig_flex_nz}(a). 
Longer chains experience a short-ranged repulsion in the vicinity of the charged wall 
($z \lesssim d$) due to {\it chain-entropy} effects.
 \footnote[2]{
 The chain-entropy effect here is merely due to the much lower number of 
 available conformations in the adsorbed state. It has to be distinguished from that
 previously discussed in \ref{sec:PE_ads_entropy}.
 }
%

When {\it image forces} come into play, (partial) monomer desorption 
sets in, whose strength increases with growing chain length $N_m$. 
This feature is due to the repulsive image-chain interaction that scales like $N_m^2$, 
whereas the attractive Wigner crystal correlations\footnote[3]{
 When charged polymers are adsorbed on the surface, they also tend to build 
 a Wigner crystal due to the strong mutual Coulomb inter-chain repulsion.
 The higher the chain length $N_m$ (i.e., the length of the chain) the stronger the effect.
 At prescribed reduced surface charge density $\sigma_s$ and monomer concentration,
 this leads to a 2D plasma term (i.e., interchain repulsion reduced by thermal energy) 
 that roughly varies like $N_m^{3/2}$, as is the case for point-like multivalent ions.
 }
scales only like $N_m^{3/2}$.

The fraction of charge $\sigma^*(z)$ of the fluid as a function of 
monomer-wall separation, $z$, is another interesting quantity to characterize
the adsorption behavior.
At $\Delta_{\epsilon}=0$, overcharging [as signaled by $\sigma^*(z)>1$] occurs as
soon as chains are longer than dimers, see figure 4(a) in \cite{Messina_PRE_2004}. 
In the presence of image forces, 
the strength of the overcharging is nearly identical to
that obtained without image forces at $\Delta_{\epsilon}=0$ 
(compare with figure 4(a) in Ref. \cite{Messina_PRE_err_2006}). 
Thereby, the main effect of image charges is (i) to decrease the fraction of charge
$\sigma^*(z)$  near contact ($z \lesssim 1.2a$) upon growing $N_m$ and 
(ii) to (slightly) shift the position of the maximum of $\sigma^*(z)$ to larger $z$.
%
%

\subsubsection{Rigid chains} \cite{Messina_JCP_2006}

Dimers exhibit a monotonic behavior for $n(z)$ that is similar to point-like ions. 
For longer chains there exits a small monomer depletion near the charged wall for an 
\textit{intermediate} regime of $N_m$, see figure \ref{fig.PE_rig_flex_nz}(a). 
At high enough $N_m$, $n(z)$ reveals again a monotonic behavior, 
see figure 1(a) in Ref. \cite{Messina_JCP_2006}.
This interesting effect is the result of two antagonistic {\it entropy}-driving forces, namely, (i) 
chain-entropy and (ii) polymerization induced adsorption. 
Electrostatic chain-chain correlations, whose strength grows in a non-trivial 
way with $N_m$,\footnote[4]{
 Due to the strong extension of the chain, it is no longer suitable to use 
 the point-like and/or spherical polyion picture leading to the WC term
 in $N_m^{3/2}$. 
 } 
favor also chain adsorption.
Figure \ref{fig.PE_rig_flex_nz}(a) clearly shows that the 
adsorption of rigid PEs is much stronger than that of flexible ones. 
This feature is also detectable in the snapshots, 
see figure \ref{fig.PE_rig_flex_nz}(b) and \ref{fig.PE_rig_flex_nz}(c).

Upon polarizing the interface, it is found that the degree of adsorption is considerably reduced.
Nonetheless, a comparison with the flexible case \cite{Messina_PRE_err_2006} shows that the values 
at contact at finite $\Delta_{\epsilon}$ are quite similar. 

\subsubsection{Summary}
 
To sum up, MC simulations \cite{Messina_PRE_2004,Messina_PRE_err_2006,Messina_JCP_2006} 
show us that:
\begin{itemize}
\item Without a dielectric discontinuity ($\Delta_{\epsilon}=0$), 
      {\it flexible} PE chains experience short-ranged repulsion near the charged substrate
      due to chain-entropy effects. In contrast, {\it rigid} PE chains are
      more strongly adsorbed (due to a weaker loss of chain-entropy) and, 
      when long enough, experience a purely effective attraction.
\item Image forces lower the degree of adsorption for flexible and rigid PE chains.
      However, the overcharging of the substrate by the PEs 
      is robust (irrespective of the chain flexibility) against image forces. 
\end{itemize}

\subsection{Polyelectrolyte multilayering}

PE multilayer thin films are often obtained using a so-called layer-by-layer 
deposition technique \cite{Decher_1992,Decher_1997}: 
A (say negatively) charged substrate is alternatively exposed to 
a polycation (PC) solution and a polyanion (PA) one.
This method and the resulting materials have a fantastic potential
of application in technology, e. g., biosensing \cite{Caruso_Langmuir_1998}, 
catalysis \cite{Onda_1999}, nonlinear optical devices \cite{Wu_JACS_1999},
nanoparticle coating \cite{Caruso_Science_1998}, etc. 

From the theoretical side the literature is rather poor. 
However, a few analytical works about PE multilayers
on charged planar surfaces based on different levels of approximation
are available \cite{Solis_JCP_1999,Netz_Macromol_1999b,Castelnovo_2000}.
Solis and Olvera de la Cruz considered the conditions under which the spontaneous
formation of polyelectrolyte layered structures can be induced by a charged
wall \cite{Solis_JCP_1999}.
Based on Debye-H\"uckel approximations for the electrostatic interactions,
but including some lateral correlations by the consideration of given
adsorbed PE structures,
Netz and Joanny\cite{Netz_Macromol_1999b} found a remarkable
stability of the (semi-flexible) PE multilayers supported by scaling laws.
For weakly charged flexible polyelectrolytes at high ionic
strength, qualitative agreements between theory
\cite{Castelnovo_2000}, also based on scaling laws, and experimental
observations \cite{Ladam_Langmuir_2000}
(such as the predicted thickness and net charge of the PE multilayer) were achieved.
More recently, Shafir and Andelman, using mean-field theory, pointed out
the relevant role of a specific strong short-range interaction between PAs and PCs.

A tremendous difficulty in PE multilayering is the strong electrostatic correlations
between PCs and PAs, which are hard to be satisfactorily taken into account in
(modified) mean-field theories. In this respect, computer simulations are of great help.
The first simulation model for PE multilayering was developed in \cite{Messina_Langmuir_2003}.
Later Panchagnula et al. performed similar computer simulations \cite{Dobrynin_PRL_2004}, 
where the dynamical aspect was more emphasized.
Several types of substrate geometry were considered, from spherical particles 
\cite{Messina_Langmuir_2003,Dobrynin_PRL_2004,Dobrynin_Langmuir_2005a}
to planar substrates 
\cite{Messina_macromol_2004,Dobrynin_Langmuir_2006}
via cylindrical ones \cite{Messina_JCP_2003b}.
Relevant simulation 
findings for spherical \cite{Messina_Langmuir_2003} 
and planar substrates \cite{Messina_macromol_2004} are going to be described.

\subsubsection{Polyelectrolyte multilayering at spherical substrates}

\begin{figure}[t]
\centering
\includegraphics[width = 12.0 cm]{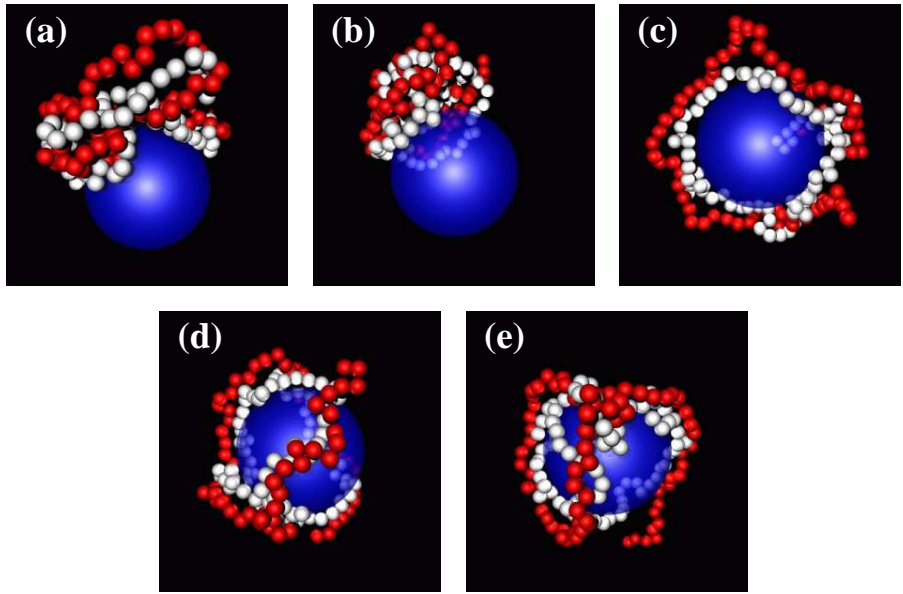}
\caption{ Typical configurations for one PC (in white) and
  one PA (in red) adsorbed onto the negatively charged colloid at
  different $\chi_{vdw}$-couplings.  (a) $\chi_{vdw}=0$ (b)
  $\chi_{vdw}=1$ (c) $\chi_{vdw}=2$ (d) $\chi_{vdw}=3$ (e)
  $\chi_{vdw}=5$.  
  Note the remarkable structural change occurring at $\chi_{vdw}=2$.
  The small univalent counterions (anions and cations) are omitted for clarity.
  }
\label{fig:2PE_bilayer_sph}
\end{figure}

\begin{figure}[]
\centering
\includegraphics[width = 12.0 cm]{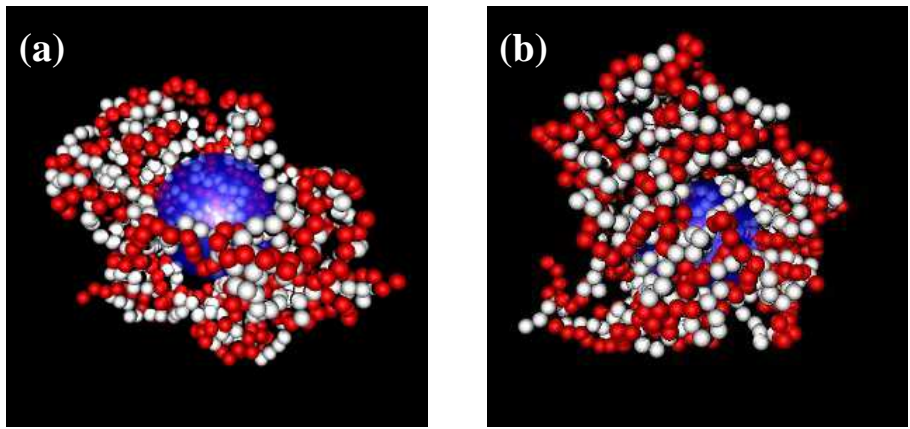}
\caption{ Typical equilibrium configurations for 12 PEs (6 PCs in
  white and 6 PAs in red) adsorbed onto the negatively charged colloid
  at different $\chi_{vdw}$-couplings.  The little counterions (anions
  and cations) are omitted for clarity.  (a) $\chi_{vdw}=0$ (b)
  $\chi_{vdw}=3$.  }
\label{fig:12PE_mlayer_sph}
\end{figure}

From the study in \cite{Messina_Langmuir_2003} concerning substrates with finite radii 
(i. e., charged spheres), one has  learned that  {\it non-electrostatic} 
forces are required to obtain (true) PE multilayers.
More precisely, by introducing a (additional) {\it short-range} van der Waals-like attraction 
(whose strength is characterized by its value at contact, $\chi_{vdw}$, in units of $k_BT$) 
between the substrate's surface and the (monomers of the) oppositely charged chains.
The PE structure results then from a complicated interplay between: 
(i) PC-PA strong attraction (favoring a collapse into a compact globular state) and
(ii) PE-substrate correlations (favoring {\it flat} adsorption and {\it wrapping} 
\footnote[1]{Note that the wrapping from the chain(s) around the colloid is peculiar
 to spherical substrates. Besides it should be reminded that wrapping is also
 governed by the repulsive interaction between the turns of a chain \cite{Grosberg_RevModPhys_2002}.}
around the sphere).
Briefly, the main findings in \cite{Messina_Langmuir_2003} are as follows:
\begin{itemize}
\item Flat {\it bilayer}-structures, involving two long oppositely charged chains, 
      set in only for large enough $\chi_{vdw}$. At low $\chi_{vdw}$, the strong driving
      PA-PC force leads to PE globular structures, see figure \ref{fig:2PE_bilayer_sph}. 
\item Stable and flat multilayers are only obtainable at large enough $\chi_{vdw}$. 
      In a purely electrostatic regime ($\chi_{vdw}=0$) PE globules are formed preventing
      a uniform coverage of the surface, see figure \ref{fig:12PE_mlayer_sph}.
\item Short chains are not suitable candidates for PE multilayers, due 
      to (i) the weaker effect of polymerization adsorption and (ii)
      reduced chain-chain correlations. 
\end{itemize}

%
%
\begin{figure}[htbp]
\centering
\includegraphics[width = 12.0 cm]{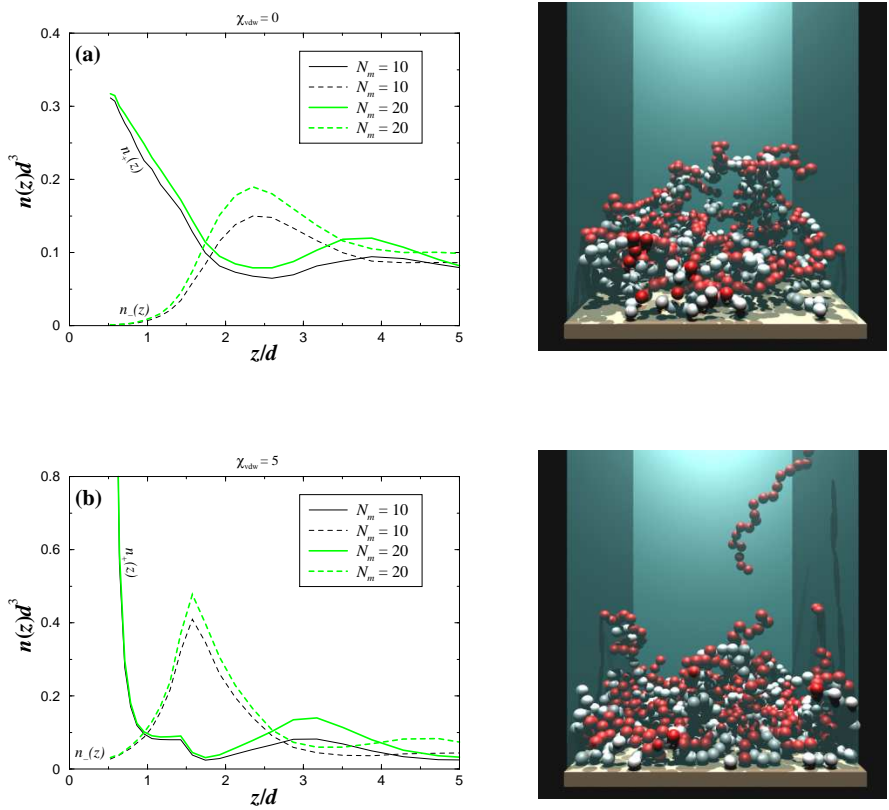}
\caption{
  Profiles of monomer density $n_{\pm}(z)$ for oppositely charged PEs adsorbed
  onto a negatively charged planar substrate.
  $\chi_{vdw}$-couplings.
  (a) $\chi_{vdw}=0$
  (b) $\chi_{vdw}=5$ \cite{Messina_macromol_2004}.
  The snapshots shown correspond to chain length $N_m=20$.
  }
\label{fig:MULTILAYER}
\end{figure}

\subsubsection{Planar substrates}

PE multilayering onto planar substrates were investigated in 
\cite{Messina_macromol_2004,Dobrynin_Langmuir_2005b,Dobrynin_Langmuir_2006}.
The zero-curvature case differs qualitatively from the spherical one. 
First the intrinsic electric field is higher in the former case
\footnote[1]{
At zero curvature we have $\psi \sim r$ in contrast to finite curvature where 
$\psi \sim 1/r$}. 
Secondly the chain-wrapping is no-longer present at zero curvature.
Consequently at given surface charge density, we expect a stronger PE-layering.    
The important results can be formulated as follows: 

\begin{itemize}
\item 
As with for spherical substrates, the relevance of short-ranged non-electrostatic
forces is also demonstrated here, see figure \ref{fig:MULTILAYER}. 
Flat multilayers can not be achieved with solely electrostatic forces.
\item 
The formation of islands (i.e., clusters of PC-PA chains) onto the substrate
are reported \cite{Messina_macromol_2004} and qualitatively confirm the experimental observations of
the early stages of PE deposition (one or two bilayers) \cite{Perez_CollSurf_2003,Harris_Langmuir_2000}.  
\end{itemize}

\section{\label{cha:2D} Confined crystalline colloids}

It is well known from solid state studies that strongly confined (i.e., quasi two-dimensional)
systems exhibit properties and a phase behavior that may drastically differ from those in the bulk
\cite{Binder_JNET_1998}.
This feature is also vivid in colloidal systems, and those materials represent ideal
model systems to analyze (experimentally as well as theoretically) and understand  
confinement effects on a mesoscopic scale corresponding to the interparticle distance.
Using external fields, a colloidal system can be prepared in a controlled way into prescribed 
equilibrium and non-equilibrium states \cite{Loewen_JPCM_2001}.
For instance, in equilibrium, solidification near interfaces (provided by a substrate or a large ``impurity'') 
can occur under thermodynamic conditions where the bulk is still  fluid 
(so-called prefreezing).
In non-equilibrium, a wall may act as a center of heterogeneous nucleation 
(favored by the excess surface-energy already offered by the wall/nucleus interface) 
and initiate crystal growth. 
Most of our experimental knowledge of freezing in a confining slit-like geometry is
based on real-space measurements of mesoscopic model systems such as charged colloidal suspensions
between glass plates \cite{Murray_PRB_1990,Neser_PRL_1997}. 

In this section, different relevant achievements in the field of
confined charged colloidal crystals are discussed.


\subsection{Two-dimensional dipolar mixtures}

Two-dimensional colloidal systems can be achieved for instance 
via sedimentation and trapping at the air/water interface 
\cite{Zahn_PRL_1997,Ebert_EPJE_2008}.
At strong external field,
all the dipolar moments are aligned in the direction of
the applied external field, leading to a purely repulsive pair interaction 
that scales like:
%
\begin{eqnarray}
\label{Eq.dipole_r3}
V_{dip}(r) \propto \frac{m_1 m_2}{r^3},
\end{eqnarray}
%
where $m_1$ and $m_2$ stand for the induced dipolar moments of the particles $1$ and $2$, respectively.
\footnote[1]{
Note that in the experimental situations, one has often to deal with 
{\it magnetic} colloidal particles (so called ferrofluids).
However {\it electric} dipoles are also realizable \cite{Yethiraj_Nature_2008}.
This being said, regardless of the nature of dipolar moment (i.e., magnetic or electric),
the same pair interaction (\ref{Eq.dipole_r3}) prevails.
Hence results on (super)magnetic particles enter also adequately in the scope of this review.   
}

Whereas the one-component situation trivially yields a triangular lattice,
the {\it binary mixture} situation provides a very rich phase behavior 
\cite{Assoud_EPL_2007}.
This feature can be conveniently exploited for potential technological applications:
optical band-gap materials (so-called photonic crystals) \cite{Blanco_Nature_2000}, 
molecular sieves \cite{Kecht_Langmuir_2004}, 
nano-filters with prescribed porosity \cite{Yan_NanoLett_2004}, etc.
There have been recent advances in this field that are going to be concisely explained here.  

\begin{figure}[]
\centering
\includegraphics[width = 12.0 cm]{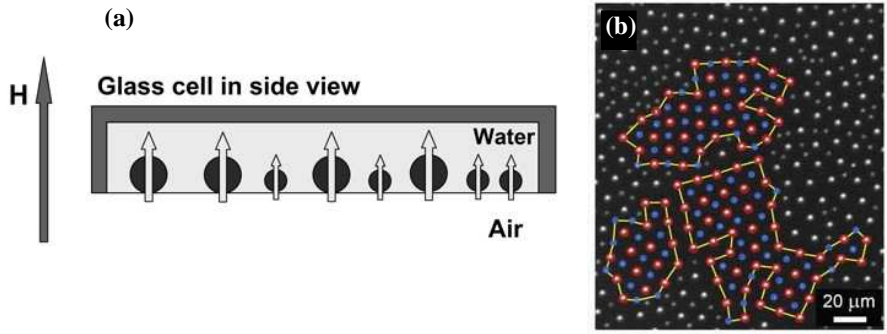}
\caption{
(a)  
 Super-paramagnetic colloidal particles confined at a water/air interface due to gravity. 
 An external magnetic field H perpendicular to the interface induces a magnetic moment
 $\vec m$ in each bead leading to a repulsive dipolar interaction, see equation (\ref{Eq.dipole_r3}).
 (b)
 Micrograph showing three touching square-latticed grains at low reduced temperature with
 a global composition $X=n_B/(n_B+n_A) \approx 45\%$ 
 [with $n_A$ and $n_B$ standing for the area density of the big and small particles, respectively] 
 and a reduced moment $m_B/m_A \approx 10\%$.
 Data taken from \cite{Ebert_EPJE_2008}.
 } 
\label{fig:2d_dipole_exp}
\end{figure}

Two dimensional binary mixtures made up of two types of dipolar particles
[(i) big particles with a large dipolar moment (species $A$) and 
(ii) small particles with a small dipolar moment (species $B$)] were investigated
experimentally \cite{Ebert_EPJE_2008}. 
The corresponding setup and a representative snapshot of the microstructure are
shown in figure \ref{fig:2d_dipole_exp}. 
A remarkable feature is the stability of the square 
phase at strong dipolar asymmetry ($m_B/m_A \approx 10 \%$).
%

\begin{figure}[]
\centering
\includegraphics[width = 14.0 cm]{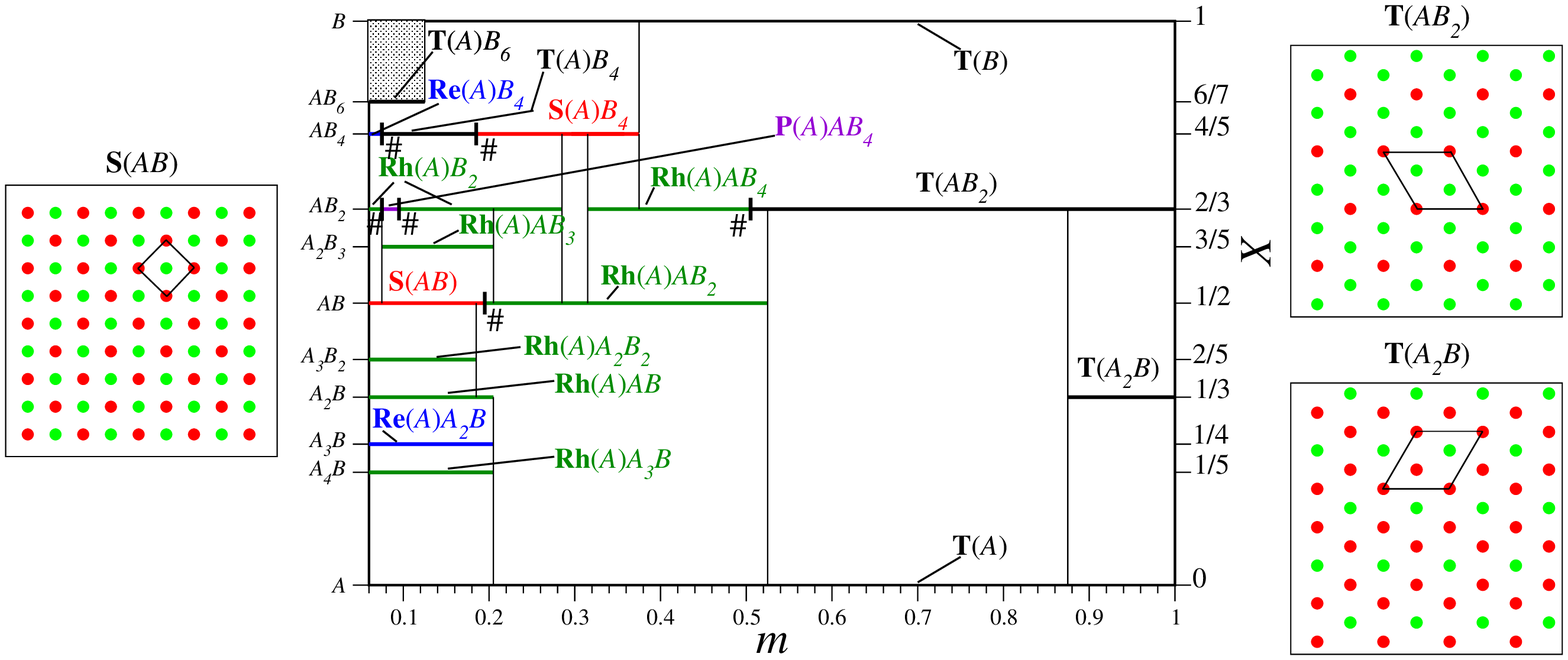}
\caption{
 The phase diagram in the $(m,X)$-plane at $T=0$.
 Three important phases are shown: ${\bf S}(AB)$, ${\bf T}(A_2B)$ and ${\bf T}(AB_2)$.
 The reader can find more details about the other structures in  \cite{Assoud_EPL_2007}.
 Data taken from \cite{Assoud_EPL_2007}.
 } 
\label{fig:2d_dipole_assoud}
\end{figure}

On the theoretical side, the phase behavior of such a binary dipolar mixture
at zero temperature was studied using lattice sums \cite{Assoud_EPL_2007}.
The relevant reduced parameters are (i) the reduced dipolar moment $m=m_B/m_A$ and
(ii) the composition $X=n_B/(n_A+n_B)$.
The resulting phase diagram is shown in figure \ref{fig:2d_dipole_assoud}.
The main results are as follows:
%
\begin{itemize}
\item 
The phase diagram qualitatively differs from that of hard disks \cite{Likos_PhilMagB_1993}.
For low dipolar asymmetry $m \gtrsim 0.5$ a stable mixture ${\bf T}(AB_2)$ sets in 
(see figure \ref{fig:2d_dipole_assoud}) 
in contrast to the case of hard-disk mixtures where {\it no} mixture is predicted 
at low size asymmetry \cite{Likos_PhilMagB_1993}. 
The stability of this phase  ${\bf T}(AB_2)$ was also reported in molecular dynamics
simulations \cite{Stirner_Langmuir_2005}. 
At even smaller dipolar asymmetry $m \gtrsim 0.88$, an additional (globally triangular)  
phase mixture ${\bf T}(A_2B)$ is stable, see figure \ref{fig:2d_dipole_assoud}.

\item 
The stability of the square phase ${\bf S}(AB)$ (see figure \ref{fig:2d_dipole_assoud})
is in good qualitative agreement with the experimental findings in \cite{Ebert_EPJE_2008},
where the dominance of the square phase is also reported 
(see figure \ref{fig:2d_dipole_exp}) as previously mentioned.

\end{itemize}

%

\subsection{Crystalline colloidal bilayers}

Crystalline bilayers made up of charged particles have been intensively studied these
last years in the soft matter colloid community 
\cite{Messina_PRL_2003,Messina_JPCM_2005a}
as well as in the solid state physics
(classical 
\cite{Goldoni_PRB_1996,Peeters_PRL_1997,Schweigert_PRB_1999,Schweigert_PRL_1999,Donko_PRL_2007} 
and non-classical electrons
\cite{Neilson_PRL_1991,Eisenstein_Nature_2004,Peeters_NanoLett_2007} ) 
and  dusty plasma communities \cite{Totsuji_PRL_1997,Zuzic_PRL_2000}.

The effective interaction between these constitutive mesoscopic macroions is neither hard-sphere like
nor purely Coulombic, but it is rather described by an intermediate screened Coulomb
[also called Yukawa or DLVO (Derjaguin-Landau-Verwey-Overbeek) \cite{DL_1941,VO_1948}]
due to the screening mediated by the additional microions present in the system.
The screening strength can be tuned by varying the microion concentration:
For colloidal systems, salt ions can be conveniently added to the aqueous suspension;  
The dusty plasma, on the other hand, consists of electrons and impurity ions.

\subsubsection{Equilibrium}
The {\it equilibrium} phase diagram at zero temperature of crystalline bilayers was 
investigated theoretically in \cite{Messina_PRL_2003}. 
The constitutive (point-like) particles interact via a Yukawa pair potential of the form 
%
\begin{eqnarray}
\label{Eq.yuk}
V_{yuk}(r)=V_0 \frac{\exp(-\kappa r)}{\kappa r}, 
\end{eqnarray}
%
where $V_0$ sets the energy scale.\footnote[1]{ Note 
 that in the ground state, i.e. at rigorously zero temperature, the value of $V_0$ is irrelevant.
 Nonetheless in experimental situations, the energy amplitude
 $V_0=Z^2 \kappa \frac{\exp(2\kappa R)}{\varepsilon (1+\kappa R)^2}$ scales like the square of the 
 charge $Z$ of the particles with a physical hard core radius $R$ reduced by the dielectric constant
 $\varepsilon$ of the solvent ($\varepsilon=1$ for the dusty plasma). For a charged colloids,
 $Z$ is typically of the order of $100-100 ~ 000$ elementary charges such that $V(r=d)$ 
 can be much larger than $k_BT$ at interparticle distance ($d$), justifying formally our
 zero-temperature calculations.} 
The choice of this potential is motivated by the experimental model systems described above.
The crystalline bilayer consists of two (identical) layers containing in total $N$ particles
in the $(x,y)$ plane. The corresponding (total) surface density $\rho$ is then given 
by $N/A$, with $A$ being the (macroscopic) layer area. 
The distance $D$, separating the two layers in the $z$-direction, is prescribed
by an (implicit) external potential confining the system. 

%
\begin{figure}[]
\centering
\includegraphics[width = 12.0 cm]{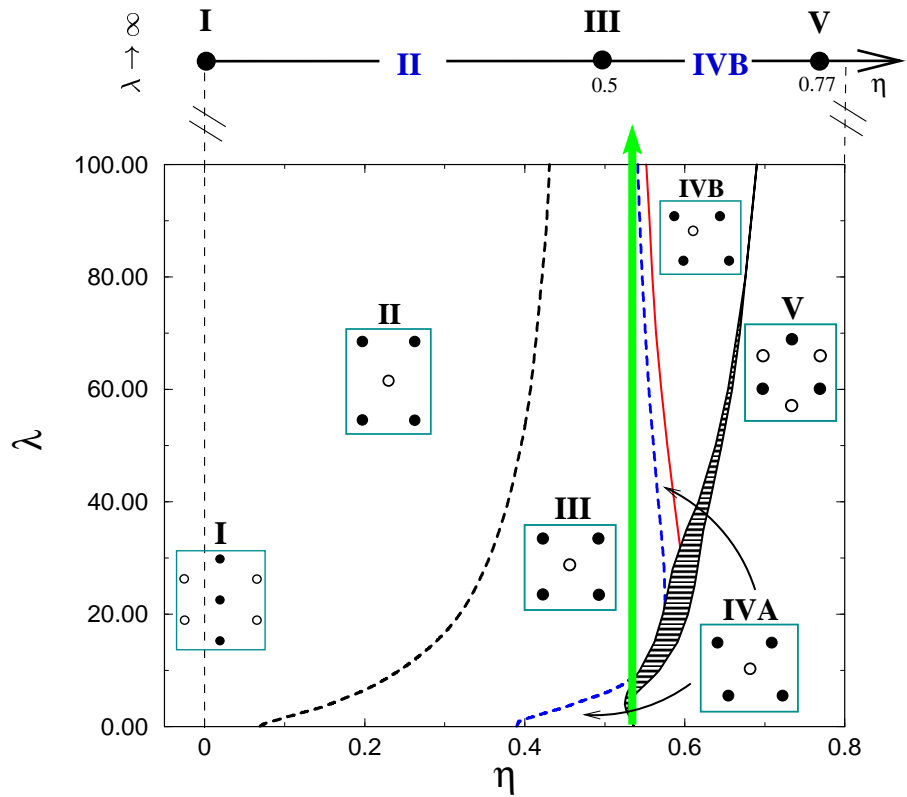}
\caption{
  The hard sphere limit $\lambda \rightarrow \infty$ is sketched on top.
  The dashed (solid) lines denote continuous (discontinuous) transitions. 
  The filled region corresponds to the coexistence domain of phases IV and V. 
  The vertical arrow indicates the \textit{double} reentrant behavior of phase IVA.
  The insets show the lattice geometries,
  where the filled (open) circles correspond to the lower (upper) layer.  
  } 
\label{fig:pd_bl_yukawa}
\end{figure}
%

The zero-temperature phase behavior is fully determined by two dimensionless parameters, 
namely the reduced layer density, $\eta=\rho D^2/2$, and the reduced screening strength, 
$\lambda=\kappa D$.
Using a straightforward lattice sum technique, the phase diagram was calculated for arbitrary
$\lambda$  and $\eta$, see figure \ref{fig:pd_bl_yukawa}.\footnote[2]{Note that 
  the ground state at vanishing screening $\lambda \to 0$ corresponds always 
  to bilyaers. Indeed, two equally charged walls  do {\it not} generate
  anay electric field within the slit, and consequently they do not alter the stable 
  Wigner crystal structure obtained at any other surface charge density 
  (including neutral walls). Thereby, if one considers the special case of two
  walls corresponding to neutralizing backgrounds, the ground state is always a bilayer.
  At finite screening $\lambda \neq 0$, however, the situation is more complicated, and multilayers
  (i.e., beyond bilayers) are stable at high enough density $\eta$. 
 } 
The most interesting findings \cite{Messina_PRL_2003} are as follows: 

\begin{itemize}
\item 
Whereas the two known extreme limits of zero 
\cite{Schweigert_PRL_1999,Schweigert_PRB_1999,Goldoni_PRB_1996} 
and infinite
\cite{Pieransky_PRL_1983,Schmidt_PRL_1996,Schmidt_PRE_1997} 
screening strength $\lambda$ are recovered by lattice sum calculations \cite{Messina_PRL_2003}, 
it is demonstrated that, at intermediate $\lambda$, 
the phase behavior is strikingly different from a simple interpolation between these two limits.
First, there is a first-order coexistence between two different staggered rhombic lattices
(IVA and IVB in figure \ref{fig:pd_bl_yukawa}) differing in their relative shift of the two unit cells.
Second, the staggered rhombic phase IVA exhibits a novel
reentrant effect for fixed density and varied screening length, see figure \ref{fig:pd_bl_yukawa}. 
Depending on the density, the reentrant transition can proceed via a staggered square III or a
staggered triangular solid V including even a {\it double reentrant transition} 
of the rhombic phase IVA, see figure \ref{fig:pd_bl_yukawa}.
%
\item
A comparative study \cite{Messina_JPCM_2005a} of the phase behavior of highly charged colloidal spheres 
in a confined wedge geometry reveals semi-quantitative agreement between 
theory and experiment.  
\end{itemize}

\subsubsection{Non-equilibrium}
The {\it non-equilibrium} case\footnote[1]{
The starting unsheared configuration corresponds to a staggered square lattice
with a reduced surface particle density $\eta=0.24$ and a reduced screening strength
$\lambda=2.5$.
Two walls are present to ensure the confinement. To this end, 
screened Coulomb and short-ranged (of the Lennard-Jones type) 
repulsive potentials were tested, and it was found that our results
are marginally sensitive to the choice of the repulsive wall-particle interaction. } 
at finite temperature as driven by a 
linear shear flow has been addressed in \cite{Messina_PRE_2006,Messina_JPCM_2005b}.
The steady state developed under shear as well as the relaxation back to equilibrium after
cessation of shear were analyzed with the help of non-equilibrium Brownian dynamics.
The pertinent results are:  
\begin{itemize}
\item 
For increasing shear rates, the following steady states are reported:
First, up to a threshold of the shear rate, there is a static solid which is
elastically sheared. 
Then, at higher shear rates the crystalline bilayer melts, and even higher shear rates
lead to a reentrant solid stratified in the shear direction.
\item 
After instantaneous cessation of shear, a nonmonotonic behavior of the typical relaxation time 
is found. In particular, application of high shear rates accelerates the relaxation back to equilibrium
since shear-induced ordering facilitates the growth of the equilibrium crystal.
\item
The orientation of a crystalline bilayer can be tuned at wish upon applying a (strong) 
shear rate in the desired direction and subsequently letting the system relax.
 
\end{itemize}



\section{\label{cha:conclu} Conclusions}

Various electrostatic effects in soft matter have been discussed.
Generally speaking, charged systems are fascinating because 
they simultaneously involve short-ranged excluded volume effects 
(as soon as the latter are properly taken into account)
already present in neutral systems, and additionally the long-ranged
Coulomb interaction. The latter constitutes a formidable theoretical challenge. 

In terms of similarities with classical solid state physics and (elementary) quantum chemistry,
two striking analogies were identified: 
(i)  The overcharging occurring at a sphere reduces to the old Thomson problem;
(ii) The ground state of two spherical macroions is ionized, with the degree of ionization 
     (and therefore the attraction strength) growing with the {\it difference in surface charge density} 
     between the two macroions. This behavior is highly reminiscent of the (molecular) ionic bonding 
     \cite{Pauling_Electronegtaivity_1939}
     where the {\it difference in electronegativity} between the two atoms governs its stability. 

Excluded volume effects are equally important to fully understand 
phenomena like overcharging (i.e., surface charge reversal) 
and surface charge amplification. For overcharging, the counterion layer  
can reach a high ordering when the local packing fraction is raised, by simply
increasing the size of the adsorbed counterions. 

Image forces turn out to be systematically short-ranged.
Their effects are only vivid close to the substrate at distances corresponding roughly to 
the linear size of the microions\footnote[1]{
The behavior might be less clear for highly charged spherical macroions as adsorbate.} 
(counterions and/or charged monomers).      
%
As far as the adsorption of polyelectrolytes is concerned, there are
two important driving forces that act concomitantly: 
(i) The polymerization-induced adsorption that works like the principle of counterion
release (so {\it entropy} based) and (ii) purely electrostatic lateral correlations
(reminiscent of the classical Wigner crystal).

Confined colloidal crystals seem to be now pretty well understood up to bilayers.
There is presently  some experimental \cite{RamiroManzano_PRE_2007,Fontecha_PRE_2007} 
and simulational \cite{Fortini_JPCM_2006} evidence that, upon increasing the 
projected surface particle density, the transition 
from two-layer to three-layer structures involve four (and even more) layered crystalline structures.
This is a problem that needs an urgent and clear understanding.

On a more ``material/engineering'' level, multilayered
structures can apparently also be experimentally obtained 
by combining oppositely charged colloids/micelles \cite{Caruso_AdvMater_2007},
instead of polyelectrolytes.
To explore this new field, a considerable theoretical effort would be needed to identify the 
parameters phase space 
(such as salt concentration, charges of the colloids and the substrates, particle size etc.) 
allowing the onset of such structures without strong clustering.    
   
\ack

The author is grateful to C. Holm, K. Kremer, H. L\"owen, and M. Lozada-Cassou
who are closely involved in jointly authored publications covered in this review.


\section*{References}


\end{document}